\documentclass[submission,copyright,creativecommons]{eptcs}
\providecommand{\event}{MARS 2020}
\usepackage{xspace}
\usepackage{multicol,multirow}
\usepackage[table]{xcolor}
\usepackage{graphicx}
\usepackage{amsfonts,amsmath,amssymb}
\usepackage[inline]{enumitem}
\setitemize{noitemsep,topsep=0pt,parsep=0pt,partopsep=0pt,leftmargin=0.7cm}
\setenumerate{noitemsep,topsep=0pt,parsep=0pt,partopsep=0pt,leftmargin=0.7cm}
\usepackage{awn,awn_keywords}
\usepackage{tikz}
\usepackage{tabu}
\usepackage{longtable}
\RequirePackage[normalem]{ulem}

\newcounter{counterfix}
\makeatletter
\long\def\algocf@caption@algo#1[#2]#3{%
  \ifthenelse{\equal{\algocf@algocfref}{\relax}}{}{\algocf@captionref}%
  \@ifundefined{hyper@refstepcounter}{\relax}{
    \ifthenelse{\equal{\algocf@algocfref}{\relax}}{\renewcommand{\theHalgocf}{\thecounterfix.\thealgocf}}{
      \renewcommand{\theHalgocf}{\thecounterfix.\algocf@algocfref}}
    \hyper@refstepcounter{algocf}
  }%
   \algocf@latexcaption{#1}[{#2}]{{#3}}
}%
\makeatother

\usepackage{listings}
\lstdefinestyle{CStyle}{
    backgroundcolor=\color{backgroundColour},   
    commentstyle=\color{mGreen},
    keywordstyle=\color{magenta},
    numberstyle=\tiny\color{mGray},
    stringstyle=\color{mPurple},
    basicstyle=\footnotesize,
    breakatwhitespace=false,         
    breaklines=true,                 
    captionpos=b,                    
    keepspaces=true,                 
    numbers=left,                    
    numbersep=5pt,                  
    showspaces=false,                
    showstringspaces=false,
    showtabs=false,                  
    tabsize=2,
    language=C
}

\definecolor{darkgreen}{rgb}{0,0.55,0}
\definecolor{darkred}{rgb}{0.8,0,0}

\newcommand{\IF}{\texttt{\ \ \bf if\hspace{13.9pt} }}
\newcommand{\THEN}{\texttt{\ \ \bf then }}
\newcommand{\ELSE}{\texttt{\ \ \bf else\hspace{3.7pt} }}
\newcommand{\ELIF}{\texttt{\ \ \bf elsif\hspace{1.9pt} }}
\newcommand{\LET}{\texttt{\ \ \bf let\ \ }}
\newcommand{\IN}{\texttt{\ \ \bf in\hspace{2.6pt}\ \ }}
\newcommand{\LETand}{\texttt{\ \ \bf let\hspace{6.1pt}\ \ }}
\newcommand{\INand}{\texttt{\ \ \bf in\hspace{8.6pt}\ \ }}
\newcommand{\AND}{\texttt{\ \ \bf and\ \ }}

\newcommand{\TRUE}{\textnormal{\textit{True}}}
\newcommand{\FALSE}{\textnormal{\textit{False}}}

\newcommand{\pred}[1]{\textnormal{\textbf{#1}}}

\SetCommentSty{mycommfont}

\renewcommand{\algorithmcfname}{Process}

\SetKwProg{GUARD}{[}{]}{}
\newcommand{\procGUARD}[2]{\GUARD{\ensuremath{\hspace{1pt} #1 \hspace{1pt}}}{#2}}
\SetKwProgNoLine{GGUARD}{[}{]\hphantom{.}}{}
\newcommand{\procGGUARD}[2]{\GGUARD{\ensuremath{\hspace{1pt} #1 \hspace{1pt}}}{#2}}
\SetKwProg{ASSN}{\color{brown}$\llbracket$}{\color{brown}$\rrbracket$}{}
\newcommand{\procASSN}[2]{\ASSN{\ensuremath{\hspace{1pt} \color{brown} #1 \hspace{1pt}}}{#2}}
\SetKw{CHOICE}{+}
\newcommand{\procCHOICE}{\CHOICE\\}
\SetKwProg{RECV}{receive(}{).}{}
\newcommand{\procRECV}[2]{\RECV{\ensuremath{\hspace{.1em}#1\hspace{.1em}}}{#2}}
\SetKwProgNoLine{RRECV}{receive(}{).}{}
\newcommand{\procRRECV}[2]{\RRECV{\ensuremath{\hspace{.1em}#1\hspace{.1em}}}{#2} \hspace{1pt}}

\SetKwProg{SEND}{send(}{).}{}

\SetKwProgNoLine{SSEND}{send(}{).}{}
\newcommand{\procSSEND}[2]{\SSEND{\ensuremath{\hspace{.1em}#1\hspace{.1em}}}{#2} \hspace{1pt}}

\SetKwProg{BC}{broadcast(}{).}{}
\newcommand{\procBC}[2]{\BC{\ensuremath{\hspace{.1em}#1\hspace{.1em}}}{#2}}
\newcommand{\procDEFN}[2]{\texttt{#1}\hspace{.1em}(\ensuremath{#2})}
\SetKwInput{OLSRDEF}{OLSR}
\SetKwInput{updateDEF}{UPDATE\_INFO}
\SetKwInput{HELLODEF}{PROCESS\_HELLO}
\SetKwInput{TCDEF}{PROCESS\_TC}
\SetKwInput{FORWARDDEF}{FORWARD\_TC}
\SetKwInput{QUEUEDEF}{QUEUE}

\newcommand{\var}[1]{\textnormal{\texttt{#1}}}

\newcommand{\typeIP}{\textnormal{\texttt{IP}}}
\newcommand{\typeTIME}{\textnormal{\texttt{TIME}}}
\newcommand{\typePOW}[1]{\ensuremath{\mathcal{P}}(\ensuremath{#1})}
\newcommand{\typeSTATUS}{\textnormal{\texttt{STATUS}}}
\newcommand{\typeMPR}{\textnormal{\texttt{MPR}}}
\newcommand{\typeSQN}{\textnormal{\texttt{SQN}}}
\newcommand{\typeMETRIC}{\textnormal{\texttt{METRIC}}}
\newcommand{\typeMsg}{\textnormal{\texttt{MESSAGE}}}
\newcommand{\typeMSG}{\textnormal{\texttt{MSG}}}
\newcommand{\typeList}[1]{\textnormal{\texttt{[#1]}}}

\newcommand{\typeL}{\textnormal{\texttt{L}}}
\newcommand{\typeNT}{\textnormal{\texttt{N2}}}
\newcommand{\typeP}{\textnormal{\texttt{P}}}
\newcommand{\typeAR}{\textnormal{\texttt{AR}}}
\newcommand{\typeTR}{\textnormal{\texttt{TR}}}
\newcommand{\typeR}{\textnormal{\texttt{R}}}
\newcommand{\typeRX}{\textnormal{\texttt{RX}}}

\newcommand{\tawn}{T-AWN\xspace}
\newcommand{\tawnopt}{(T-)AWN\xspace}
\newcommand{\awn}{AWN\xspace}
\newcommand{\tTIME}{\keyw{TIME}\xspace}

\newcommand{\now}{\keyw{now}\xspace}

\newcommand{\updl}{\texttt{[}}
\newcommand{\updr}{\texttt{]}}
\newcommand{\setl}{\texttt{\{}}
\newcommand{\setr}{\texttt{\}}}
\renewcommand{\var}{\keyw}
\newcommand{\llbracket}{\ensuremath{\textbf{[\![}}}
\newcommand{\rrbracket}{\ensuremath{\textbf{]\!]}}}

\newcommand{\typeNAT}{\NN}
\newcommand{\NN}{
    \ensuremath{%
        \mathop{\rm I\mkern-2.5mu N}%
        \nolimits%
    }%
}
\newcommand{\typeINT}{\ZZ}
\newcommand{\ZZ}{%
    \ensuremath{%
        \mathop{\rm \hspace*{0.3em}Z\hspace{-0.875em}Z\hspace{0.35em}}%
        \nolimits%
    }%
}
\newcommand{\typeBOOL}{\BB}
\newcommand{\BB}{%
    \ensuremath{%
        \mathop{{\rm I}\!{\rm B}}%
        \nolimits%
    }%
}

\title{Formalising the Optimised Link State Routing Protocol}
\def\titlerunning{Formalising OLSR}
\author{Ryan Barry
\institute{School of Computer Science and Engineering\\
University of New South Wales, Sydney, Australia}%
\email{ryan.barry@unsw.edu.au}
\and
Rob van Glabbeek
\institute{Data61, CSIRO, Sydney, Australia}
\institute{School of Computer Science and Engineering\\
University of New South Wales, Sydney, Australia}%
\email{rvg@cs.stanford.edu}
\and
Peter H\"ofner
\institute{Research School of Computer Science\\ The Australian National University, Australia}
\institute{Data61, CSIRO, Sydney, Australia}
\institute{School of Computer Science and Engineering\\ University of New South Wales, Sydney, Australia}%
\email{Peter.Hoefner@anu.edu.au}
}
\def\authorrunning{R. Barry, R.J. van Glabbeek \& P. H\"ofner}
\begin{document}
\SetAlgorithmName{Process}{Process}{List of Processes}

\maketitle

\begin{abstract}
Routing protocol specifications are traditionally written in plain English.
Often this yields ambiguities, inaccuracies or even contradictions.
Formal methods techniques, such as process algebras, avoid these problems, thus leading to more precise and verifiable descriptions of protocols.
In this paper we use the timed process algebra T-AWN for modelling the Optimised Link State Routing protocol (OLSR) version~2.
\end{abstract}

\section{Introduction}

Wireless Mesh Networks (WMNs) are a promising technology, having seen recent successes in the area of wireless communication. These multi-hop networks are designed to operate in a decentralised manner, with the responsibility of route discovery and packet forwarding being placed upon the nodes comprising a network. This necessitates the exchange of control messages in order to determine routes to available destinations.

A subset of WMNs known as Mobile Ad hoc Networks (MANETs) have received considerable attention for their use in vehicular communication\footnote{Here, MANETs are often called Vehicular Ad hoc Networks (VANETs).}, disaster relief and other emerging fields. Nodes in a MANET are distinguished by their high mobility when compared to other types of mesh networks. As a consequence of link breakages and fluctuations in signal quality, routes through these networks are subject to frequent changes. To further complicate matters, MANETs often require restrictions on bandwidth and power consumption due to the limitations of physical devices.

To address the challenges associated with ad hoc routing, a number of protocols specifically tailored to MANETs have been proposed in the literature. Of these protocols, recent research and development efforts by the Internet Engineering Task Force (IETF) have primarily targeted the Optimised Link State Routing protocol (OLSR) \cite{rfc3626}. Originally specified in 2003, the protocol received numerous design modifications over the next decade which culminated in OLSR version 2 (OLSRv2) \cite{rfc7181}.

Despite its status as an IETF proposed standard, analyses of OLSRv2 have thus far been limited to simulations and testbed experiments. These techniques, although instrumental to the development of MANET protocols, cannot guarantee certain desirable properties, or the absence of certain undesirable properties, in a system. Given that MANETs are increasingly deployed in safety-critical applications, stronger correctness guarantees ought to be made about the protocol.

Due to the ambiguous nature of natural languages, protocol specifications are often filled with ambiguities and contradictions that give rise to conflicting interpretations. To avoid such issues, we model OLSRv2 in a formalism known as the Timed Algebra for Wireless Networks (T-AWN) \cite{ESOP16,tawn_tr}.

T-AWN, and its untimed fragment AWN \cite{ESOP12}, have been used previously to formalise the routing protocol AODV \cite{rfc3561}---see \cite{DIST16,tawn_tr}. This turned out to be a solid basis for analysing the protocol by means of model checking~\cite{TACAS12}. Additionally, based on this formalisation, crucial correctness properties of the protocol, including loop freedom, route correctness and route discovery, have been proven \cite{DIST16,BGH14b,tawn_tr} or disproven \cite{GHTP13,TR13,tawn_tr}, manually and with the support of interactive proof assistants. We expect the present formalisation of OLSRv2 to be equally useful for establishing correctness properties.

\section{The Optimised Link State Routing Protocol}
The Optimised Link State Routing protocol version 2 (OLSRv2) \cite{rfc7181} is a link-state protocol tailored specifically to MANETs. In contrast to its 2003 counterpart, OLSRv1 \cite{rfc3626}, OLSRv2's offering of improved security, flexibility and scalability has solidified its status as the sole IETF ``proposed standard'' among MANET routing protocols.

\noindent Before describing the protocol's operation, we clarify some basic terminology.
\begin{description}
	\item [Symmetric path:] A sequence of nodes in a graph $\dval{ip}_0...\dval{ip}_n$ is a symmetric path iff for each pair of nodes $(\dval{ip}_i,\dval{ip}_{i+1})$ there exists an edge from $\dval{ip}_i$ to $\dval{ip}_{i+1}$ and vice-versa.
	\item[Symmetric n-hop neighbour:] A node $\dval{ip}$ is a symmetric $n$-hop neighbour of another node $\dval{ip}'$ iff $\dval{ip} \neq \dval{ip}'$ and there exists a symmetric path of exactly $n$ edges between $\dval{ip}$ and $\dval{ip}'$. 
		(So, a symmetric $n$-hop neighbour is also a symmetric $n{+}2$-hop neighbour.)
\end{description}

OLSRv2 is based on traditional link-state routing,
and so inherits its basic characteristics. Each node maintains a graph representing the network topology, with each edge representing a pair of symmetric 1-hop neighbours. Optimal routes to all reachable destinations are then determined by applying some shortest path algorithm to this graph. Rather than waiting for data packets to arrive, the topology graph is proactively assembled by exchanging link-state information with neighbouring routers in the network, through broadcasts. 
These advertisements are scheduled periodically by each node using local timers.

Motivated by the strict limits on power and bandwidth consumption in MANETs, the designers of OLSR pioneered a number of optimisations to control traffic generation. The most significant optimisations come in the form of flooding reduction, where the number of broadcasts is reduced, and topology reduction, where the number of advertised links is minimised. Each router designates a subset of its symmetric 1-hop neighbours with one or both of these tasks in a process known as \emph{multipoint relay} (MPR) selection.

\begin{description}
	\item[MPR:] A router (node) $\dval{ip}$ is an MPR of another router $\dval{ip}'$ iff $\dval{ip}$ has been designated the task of flooding reduction or topology reduction on behalf of $\dval{ip}'$.
	\item[MPR selector:] A router $\dval{ip}'$ is an MPR selector of another router $\dval{ip}$ iff $\dval{ip}$ is an MPR of $\dval{ip}'$.
\end{description}

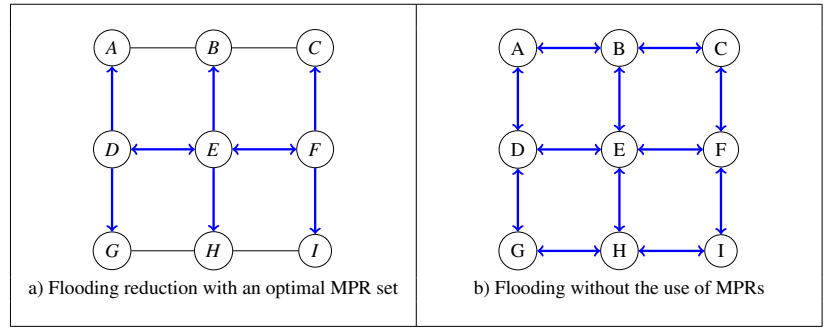
\begin{figure}
  \centering
  \scalebox{0.75}{
  \scalebox{0.9}{
\begin{tabu} to \textwidth{|X[c]|X[c]|}
    \hline
    & \\
\begin{tikzpicture}
    \node[shape=circle,draw=black] (A) at (0,4) {$A$};
    \node[shape=circle,draw=black] (B) at (2,4) {$B$};
    \node[shape=circle,draw=black] (C) at (4,4) {$C$};
    \node[shape=circle,draw=black] (D) at (0,2) {$D$};
    \node[shape=circle,draw=black] (E) at (2,2) {$E$};
    \node[shape=circle,draw=black] (F) at (4,2) {$F$};
    \node[shape=circle,draw=black] (G) at (0,0) {$G$};
    \node[shape=circle,draw=black] (H) at (2,0) {$H$};
    \node[shape=circle,draw=black] (I) at (4,0) {$I$};
    \path [-] (A) edge node[] {} (B);
    \path [-] (B) edge node[] {} (C);
    \path [<->,line width=1.3pt,blue] (D) edge node[] {} (E);
    \path [<->,line width=1.3pt,blue] (E) edge node[] {} (F);
    \path [-] (G) edge node[] {} (H);
    \path [-] (H) edge node[] {} (I);
    \path [<-,line width=1.3pt,blue] (A) edge node[] {} (D);
    \path [->,line width=1.3pt,blue] (D) edge node[] {} (G);
    \path [<-,line width=1.3pt,blue] (B) edge node[] {} (E);
    \path [->,line width=1.3pt,blue] (E) edge node[] {} (H);
    \path [<-,line width=1.3pt,blue] (C) edge node[] {} (F);
    \path [->,line width=1.3pt,blue] (F) edge node[] {} (I);
\end{tikzpicture}
&
\begin{tikzpicture}
    \node[shape=circle,draw=black] (A) at (0,4) {A};
    \node[shape=circle,draw=black] (B) at (2,4) {B};
    \node[shape=circle,draw=black] (C) at (4,4) {C};
    \node[shape=circle,draw=black] (D) at (0,2) {D};
    \node[shape=circle,draw=black] (E) at (2,2) {E};
    \node[shape=circle,draw=black] (F) at (4,2) {F};
    \node[shape=circle,draw=black] (G) at (0,0) {G};
    \node[shape=circle,draw=black] (H) at (2,0) {H};
    \node[shape=circle,draw=black] (I) at (4,0) {I};
    \path [<->,line width=1.3pt,blue] (A) edge node[] {} (B);
    \path [<->,line width=1.3pt,blue] (B) edge node[] {} (C);
    \path [<->,line width=1.3pt,blue] (D) edge node[] {} (E);
    \path [<->,line width=1.3pt,blue] (E) edge node[] {} (F);
    \path [<->,line width=1.3pt,blue] (G) edge node[] {} (H);
    \path [<->,line width=1.3pt,blue] (H) edge node[] {} (I);
    \path [<->,line width=1.3pt,blue] (A) edge node[] {} (D);
    \path [<->,line width=1.3pt,blue] (D) edge node[] {} (G);
    \path [<->,line width=1.3pt,blue] (B) edge node[] {} (E);
    \path [<->,line width=1.3pt,blue] (E) edge node[] {} (H);
    \path [<->,line width=1.3pt,blue] (C) edge node[] {} (F);
    \path [<->,line width=1.3pt,blue] (F) edge node[] {} (I);
\end{tikzpicture}
\\
a) Flooding reduction with an optimal MPR set & b) Flooding without the use of MPRs \\
 & \\
    \hline
\end{tabu}
}
  }
  \caption{Flooding reduction}
  \label{fig:flooding}
\end{figure}

Routers in OLSRv2 maintain two sets of MPRs for flooding reduction and topology reduction, 
respectively. Flooding MPRs are responsible for forwarding link advertisements received from their
flooding MPR selectors. To ensure that all routers receive link advertisements, each router must
guarantee that all of its symmetric 2-hop neighbours are symmetric 1-hop neighbours of a flooding
MPR, or symmetric 1-hop neighbours themselves. 
\pagebreak[3]
\autoref{fig:flooding} demonstrates this flooding
reduction in practice. Just three broadcasts are needed to disseminate $E$'s link advertisements when using optimal flooding MPR sets, whereas all nine nodes must perform a broadcast when flooding reduction is disabled.
By contrast, routing MPRs are responsible for advertising links between themselves and their routing MPR selectors. Routing MPRs are chosen such that all symmetric 2-hop neighbours are accessible via a routing MPR in a minimal distance 1-hop or 2-hop route. In theory, the links advertised by routing MPRs are sufficient for all routers to construct shortest paths through the network. 

\begin{figure}[!ht]
  \centering
  \scalebox{0.8}{
  \begin{footnotesize}
\begin{tabu} to \textwidth{|X[l]|X[l]|}
    \hline
     & \\[-0.5em]
\centering\scalebox{1}{
\begin{tikzpicture}
    \node[shape=circle,draw=black] (A) at (0,0) [label=$\{\ \}$][label=below:$\{\ \}$]{A};
    \node[shape=circle,draw=black] (B) at (3,0) [label=$\{\ \}$][label=below:$\{\ \}$]{B};
    \node[shape=circle,draw=black] (C) at (6,0) [label=$\{\ \}$][label=below:$\{\ \}$]{C};
    \path [-] (A) edge node[] {} (B);
    \path [-] (B) edge node[] {} (C);
\end{tikzpicture}
}
&
\centering\scalebox{1}{
\begin{tikzpicture}
    \node[shape=circle,draw=black] (A) at (0,0) [label=$\{\ \}$][label=below:$\{\ \}$]{A};
    \node[shape=circle,draw=black] (B) at (3,0) [label={\{(A,$\rightarrow$)\}}][label=below:$\{\ \}$]{B};
    \node[shape=circle,draw=black] (C) at (6,0) [label=$\{\ \}$][label=below:$\{\ \}$]{C};
    \path [->,thick,blue] (A) edge node[above] {\tiny\{ \}} (B);
    \path [-] (B) edge node[] {} (C);
\end{tikzpicture}
}
\\
	\begin{enumerate}[label=\alph*), start=1,align=left, labelwidth=0pt, itemindent=!, labelsep=3pt, leftmargin=10pt]
    		\item The initial state of the network. Top sets contain tuples of 1-hop neighbours. Bottom sets contain tuples of symmetric 2-hop neighbours.
     \end{enumerate}
     & 
     \begin{enumerate}[label=\alph*), start=2, align=left, labelwidth=0pt, itemindent=!, labelsep=3pt, leftmargin=10pt]
     	\item A sends the first HELLO message, advertising its presence. B creates a heard tuple for A since A has not yet heard from B. 
     \end{enumerate}\\
    \hline
     & \\[-0.5em]
\centering\scalebox{1}{
\begin{tikzpicture}
    \node[shape=circle,draw=black] (A) at (0,0) [label={\{(B,$\leftrightarrow$)\}}][label=below:$\{\ \}$]{A};
    \node[shape=circle,draw=black] (B) at (3,0) [label={\{(A,$\rightarrow$)\}}][label=below:$\{\ \}$]{B};
    \node[shape=circle,draw=black] (C) at (6,0) [label={\{(B,$\rightarrow$)\}}][label=below:$\{\ \}$]{C};
    \path [<-,thick,blue] (A) edge node[above] {\tiny\{(A,$\rightarrow$)\}} (B);
    \path [->,thick,blue] (B) edge node[above] {\tiny\{(A,$\rightarrow$)\}} (C);
\end{tikzpicture}
}
&
\centering\scalebox{1}{
\begin{tikzpicture}
    \node[shape=circle,draw=black] (A) at (0,0) [label={\{(B,$\leftrightarrow$)\}}][label=below:$\{\ \}$]{A};
    \node[shape=circle,draw=black] (B) at (3,0) [label={\{(A,$\rightarrow$),(C,$\leftrightarrow$)\}}][label=below:$\{\ \}$]{B};
    \node[shape=circle,draw=black] (C) at (6,0) [label={\{(B,$\rightarrow$)\}}][label=below:$\{\ \}$]{C};
    \path [-] (A) edge node[] {} (B);
    \path [<-,thick,blue] (B) edge node[above] {\tiny\{(B,$\rightarrow$)\}} (C);
\end{tikzpicture}
}
\\
	\begin{enumerate}[label=\alph*), start=3,align=left, labelwidth=0pt, itemindent=!, labelsep=3pt, leftmargin=10pt]
    		\item B sends a HELLO message. A creates a symmetric tuple for B since B has already heard from A. C creates a heard tuple for B.
	\end{enumerate}
	&
	\begin{enumerate}[label=\alph*), start=4,align=left, labelwidth=0pt, itemindent=!, labelsep=3pt, leftmargin=10pt]
    		\item C sends a HELLO message. B creates a symmetric tuple for C.
    \end{enumerate}\\
    \hline
     & \\[-0.5em]
\centering\scalebox{1}{
\begin{tikzpicture}
    \node[shape=circle,draw=black] (A) at (0,0) [label={\{(B,$\leftrightarrow$)\}}][label=below:$\{\ \}$]{A};
    \node[shape=circle,draw=black] (B) at (3,0) [label={\{(A,$\leftrightarrow$),(C,$\leftrightarrow$)\}}][label=below:$\{\ \}$]{B};
    \node[shape=circle,draw=black] (C) at (6,0) [label={\{(B,$\rightarrow$)\}}][label=below:$\{\ \}$]{C};
    \path [->,blue,thick] (A) edge node[above] {\tiny\{(B,$\leftrightarrow$)\}} (B);
    \path [-] (B) edge node[] {} (C);
\end{tikzpicture}
}
&
\centering\scalebox{1}{
\begin{tikzpicture}
    \node[shape=circle,draw=black] (A) at (0,0) [label={\{(B,$\leftrightarrow$)\}}][label=below:{\{(C, B)\}}]{A};
    \node[shape=circle,draw=black] (B) at (3,0) [label={\{(A,$\leftrightarrow$),(C,$\leftrightarrow$)\}}][label=below:{\{\ \}}]{B};
    \node[shape=circle,draw=black] (C) at (6,0) [label={\{(B,$\leftrightarrow$)\}}][label=below:{\{(A, B)\}}]{C};
    \path [<-,thick,blue] (A) edge node[above] {\tiny\{(A,$\leftrightarrow$),(C,$\leftrightarrow$)\}} (B);
    \path [->,thick,blue] (B) edge node[above] {\tiny\{(A,$\leftrightarrow$),(C,$\leftrightarrow$)\}} (C);
\end{tikzpicture}
}
\\
	\begin{enumerate}[label=\alph*), start=5,align=left, labelwidth=0pt, itemindent=!, labelsep=3pt, leftmargin=10pt]
    		\item After a certain amount of time has passed, A sends its second HELLO message. B creates a symmetric tuple for A.
	\end{enumerate}
    &
	\begin{enumerate}[label=\alph*), start=6,align=left, labelwidth=0pt, itemindent=!, labelsep=3pt, leftmargin=10pt]
    		\item B sends its second HELLO message. C creates a symmetric tuple for B. Since they are both accessible via B, A and C create 2-hop tuples for each other. Neighbourhood discovery completes.
	\end{enumerate}\\
    \hline
\end{tabu}
\end{footnotesize}
  }
  \caption{A simple Hello message exchange}
  \label{fig:hello}
\end{figure}
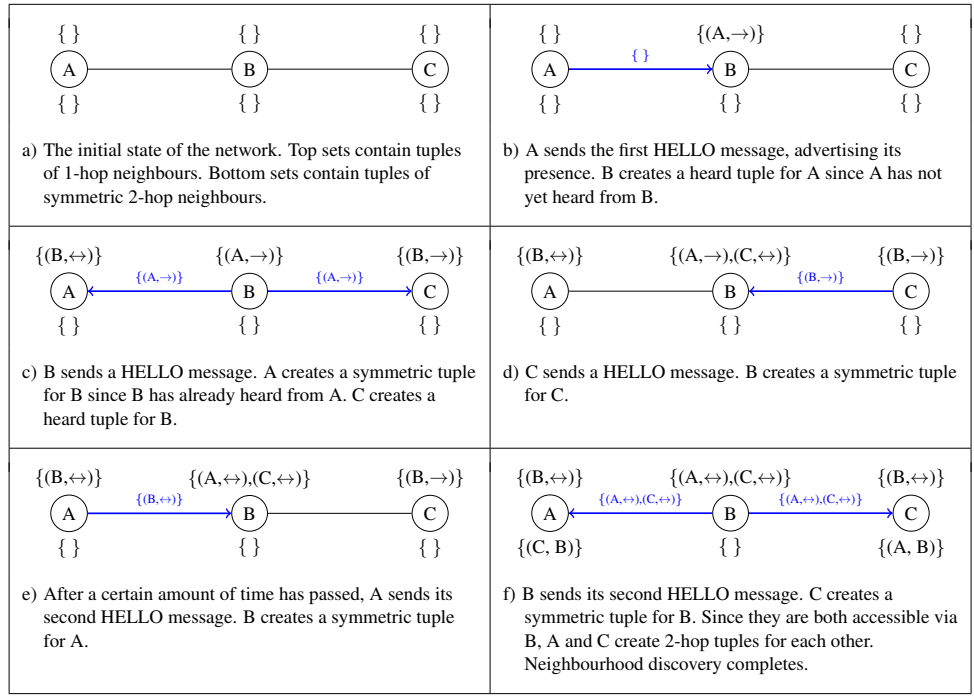

Before links between routers can be advertised, each router must identify all of its symmetric 1-hop and 2-hop neighbours. The Neighbourhood Discovery Protocol (NHDP) formalised in RFC 6130~\cite{rfc6130} is incorporated into and extended by OLSRv2 for this purpose. At a specific interval, a router will broadcast a HELLO message containing the addresses and statuses of its 1-hop neighbours. On the receiving end, the sending router is assigned a status of either symmetric, indicating a bidirectional link, or heard, indicating a unidirectional link. This status is determined by the receiver's inclusion in the HELLO message. Once symmetric links between routers are established, symmetric 2-hop neighbours can be inferred from the contents of current and future HELLO messages. All HELLO messages contain a validity time which determines when this information must be discarded, so care should be taken to avoid premature timeouts while links still exist. A simplified HELLO message exchange is detailed in \autoref{fig:hello}.

OLSRv2 extends NHDP with the inclusion of link metrics and MPR sets in its HELLO messages. Link metrics are assessed at the receiving end, and must therefore be propagated backwards to the sending router. Meanwhile, the flooding and routing MPR sets included in these messages indicate to the receiving routers whether or not they should engage in link advertisement on behalf of the sender.

After a node has identified its routing MPR selectors, it advertises all links between itself and these routers. The dissemination of this link-state information, when used in tandem with links discovered by NHDP, allows routers to establish shortest paths to all reachable destinations. The messages used to propagate advertised links are referred to as Topology Control (TC) messages. These messages are generated periodically and flooded through the network so that all reachable destinations may receive~them. 

TC messages contain a set of links to advertised routers, in addition to an advertising neighbour sequence number indicating how recent the message is. Unlike HELLO messages, which are processed but never forwarded, TC messages are flooded through the network by the flooding MPRs of each broadcasting router. When a router receives a TC message, it updates its topology graph with the advertised links provided that the advertising neighbour sequence number included in the message does not indicate out-of-date information. The message is then forwarded if it was received from a flooding MPR selector and was not forwarded in the past. Like HELLO messages, TC messages are prone to expiration if not received frequently.


\section{The Specification Language \tawn}
One of the standard tools for describing interactions, communications and synchronisations between a collection of agents, processes or network nodes is provided by process algebras. Process algebras are a family of approaches to modelling concurrent systems, as well as formally analysing said systems through algebraic laws. We choose to model OLSRv2 using \tawn~\cite{ESOP16,tawn_tr}, a \emph{timed} process algebra 
designed for wireless networks in general and routing protocols in particular.

The reason for choosing \tawn is two-fold: on the one hand, it is tailored to wireless protocols and therefore offers primitives such as {\bf broadcast};  on the other hand, it defines the protocol in pseudo-code  that is easily readable by any network or software researcher/engineer. The language itself is implementation independent. 

The timed process algebra \tawn is based on the (untimed) process algebra \awn (Algebra of Wireless Networks)~\cite{ESOP12,TR13}. \tawnopt's key operators are {\em conditional unicast}---allowing error handling in response to failed communications while abstracting from link layer implementations of the communication handling---and {\em local broadcast}---allowing a node to send messages to all its immediate neighbours as implemented by the physical and data link layer, i.e.\ to all neighbours within transmission range.

Every process algebra such as {\tawnopt} is equipped with an operational semantics \cite{ESOP12,ESOP16}: once a model has been given, its (timed) behaviour is governed by the transitions allowed by the algebra's semantics. In this paper we abstain from a formal definition of the operational semantics. Instead, we employ a correspondence between the transitions of \tawnopt processes and the execution of \emph{actions}---subexpressions as occur in Entries 3--10 of  \autoref{tb:procexpr}---identified by line numbers in protocol specifications in \tawnopt. 

We use an underlying data structure (sketched in \autoref{sec:model} and described in detail in \autoref{app:data}) with several types, variables ranging over these types, operators and predicates. First order predicate logic yields terms (or \emph{data expressions}) and formulas to denote data values and statements about them. The {\tawnopt} data structure must contain the types \tDATA, \tMSG, {\tIP} and $\pow(\tIP)$ of \emph{application data}, \emph{messages}, \emph{IP addresses}---or any other node identifiers---and \emph{sets of IP addresses}, respectively; in the case of \tawn, the data structure also features the type \tTIME of time values. The rest of the data structure is customisable for any application of \tawnopt.

In the process algebra at hand an entire network is modelled as an encapsulated parallel composition of network nodes; several processes can be executed on the same node. Nodes can only communicate with their direct neighbours, i.e.\ with nodes that are currently within transmission range. There are three different ways for nodes to perform internode communication: broadcast, unicast, or an iterative unicast/multi\-cast (called \emph{groupcast} in \tawnopt).

The \emph{process expressions} are given in \autoref{tb:procexpr}.
 \begin{table}[t]
\caption{process expressions~\cite{DIST16}}\vspace{1ex}
 \centering
{\small
  \setlength{\tabcolsep}{2.6pt}
 \begin{tabular}{|@{\ \ }l@{\ \ }|@{\ \ }p{9.5cm}|}
\hline
\rule[6.5pt]{0pt}{1pt}%
$X(\dexp{exp}_1,\ldots,\dexp{exp}_n)$& process name with arguments\\
$\p+\q$ & choice between processes $\p$ and $\q$\\
$\cond{\varphi}\p$&conditional process (if-statement)\\
$\assignment{\keyw{var}:=\dexp{exp}}\p$&assignment followed by process $\p$\\
$\broadcastP{\dexp{ms}}.\p $&broadcast of message \dexp{ms} followed by $\p$\\
$\groupcastP{\dexp{dests}}{\dexp{ms}}.\p$&iterative unicast or
  multicast to all destinations \dexp{dests}\\
$\unicastP{\dexp{dest}}{\dexp{ms}}.\p \prio \q$& unicast $\dexp{ms}$ to $\dexp{dest}$; if successful proceed with $\p$; otherwise with $\q\hspace{-2.5pt}$\\
$\send{\dexp{ms}}.\p$&synchronously transmit \dexp{ms} to  parallel process on same node\\
$\deliver{\dexp{data}}.\p$&deliver data to application layer\\
$\receive{\msg}.\p$&receive a message\\
\hline
\rule[6.5pt]{0pt}{1pt}%
$\xi,\p$         &process with valuation\\
$P\parl Q$	&parallel processes on the same node\\
\hline
\rule[6.5pt]{0pt}{1pt}%
$a\mathop{:}P\mathop{:}R$  & node $a$ running process $P$ with range $R$\\
$N\|M$		&parallel composition of nodes\\
$[N]$		&encapsulation\\
\hline
\end{tabular}}
\label{tb:procexpr}
\end{table}
A process name $X$ comes with a \emph{defining equation}\vspace{-1ex}
	\[X(\keyw{var}_1,\ldots,\keyw{var}_n) \stackrel{{\it def}}{=} \p\,,\]
where $\p$ is a process expression, and the $\keyw{var}_i$ are data variables maintained by process $X$. Furthermore, $\varphi$ is a condition, $\keyw{var}\mathop{:=}\dexp{exp}$ an assignment of a data expression \dexp{exp} to a variable \keyw{var} of the same type, \dexp{dest}, \dexp{dests}, \dexp{data} and \dexp{ms} data expressions of types {\tIP}, $\pow(\tIP)$, {\tDATA} and {\tMSG}, respectively, and $\msg$ a data variable of type \tMSG.

Given a valuation of the data variables by concrete data values, the  process $\cond{\varphi}\p$ acts as $\p$ if $\varphi$ evaluates to {\tt true}, and deadlocks if $\varphi$ evaluates to {\tt false}.\footnote{As \label{fn:undefvalues}operators we also allow \emph{partial} functions with the convention that any atomic formula containing an undefined subterm evaluates to {\tt false}.} In case $\varphi$ contains free variables that are not yet interpreted as data values, values are assigned to these variables in any way that satisfies $\varphi$, if possible. The  process $\assignment{\keyw{var}\mathop{:=}\dexp{exp}}\p$ acts as $\p$, but under an updated valuation of the data variables. The  process $\p+\q$ acts either as $\p$ or as $\q$, depending on which of the two processes is able to act at all.  In case both are able to act, the choice is non-deterministic. The process $\broadcastP{\dexp{ms}}.\p$ broadcasts (the data value bound to the expression) $\dexp{ms}$ to all nodes within transmission range, and subsequently acts as $\p$, whereas the process $\unicastP{\dexp{dest}}{\dexp{ms}}.\p \prio \q$ tries to unicast the message $\dexp{ms}$ to the destination \dexp{dest}; if successful it continues to act as $\p$ and otherwise as $\q$. The unicast is unsuccessful if the destination \dval{dest} is out of transmission range of the node $\dval{ip}$ performing the unicast. The latter models an abstraction of an acknowledgment-of-receipt mechanism that is typical for unicast communication but absent in broadcast communication, as implemented by the link layer of wireless standards such as IEEE 802.11~\cite{IEEE80211}. The process $\groupcastP{\dexp{dests}}{\dexp{ms}}.\p$ tries to transmit \dexp{ms} to all destinations $\dexp{dests}$, and proceeds as $\p$ regardless of whether any of the transmissions is successful. The process $\send{\dexp{ms}}.\p$ synchronously transmits a message to another process running on the same network node; this action can occur only when the other process is ready to receive the message. The process $\receive{\msg}.\p$ receives a message $m$  (of type \tMSG); the  value $m$ is then bound to the variable $\msg$ and the process proceeds as $\p$. The received message stems either from another node, from another process  running on the same node or from the application layer process on the local node. The latter is used to model the injection of data to the network, using the process $\receive{\newpkt{\dexp{data}}{\dexp{dip}}}$, where the function $\newpktID$ generates a message containing the application layer $\dexp{data}$ and the intended destination address $\dexp{dip}$. Data is delivered back to the application layer by \deliver{\dexp{data}}.

A state of a network node is modelled as a \emph{valuated process} given as a pair $(\xi, \p)$ of a process expression $\p$ built from the above syntax, together with a (partial) \emph{valuation} function $\xi$ that specifies values of the data variables maintained by $\p$. Finally, $P\parl Q$ denotes a parallel composition of processes $P$ and $Q$, with information piped from right to left; in our application $Q$ will be a message queue.

In the full process algebras \awn \cite{ESOP12} and \tawn \cite{ESOP16}, \emph{node expressions} $a\mathop{:}P\mathop{:}R$ are given by process expressions $P$, annotated with an \emph{address}~$a$ and a set of nodes $R$ that are within \emph{transmission range} of~$a$. A partial network is then modelled as a parallel composition of node expressions, using the operator $\|$, and a complete network is obtained by placing this composition in the scope of an encapsulation operator $[\_\!\_\,]$. The main purpose of the encapsulation operator is to prevent the receipt of messages that have never been sent by other nodes in the network---with the exception of messages $\newpkt{\dexp{data}}{\dexp{dip}}$ stemming from the application layer of a node.

When designing or formalising a protocol in \tawnopt, an engineer should not be bothered with timing aspects, except for functions and procedures that schedule tasks depending on the current time. Because of this, the only difference between the syntax of \awn and the one of \tawn is that the latter is equipped with a local timer \now{}, which is of type \tTIME.

\tawn assumes a discrete model of time, where each sequential process maintains the local variable \now holding its local clock value---an integer. Only one clock for each sequential process is employed. All (sequential) processes in a network synchronise in taking time steps, and at each time step all local clocks are incremented by one time unit.  For the rest, the variable \now behaves as any other variable maintained by a process: its value can be read when evaluating guards, thereby making progress time-dependent, and any value can be assigned to it, thereby resetting the local clock.

Before describing our \tawn-specification of OLSRv2, we want to point out two fundamental assumptions of \tawn. (i) The underlying formal semantics of \tawnopt is that any broadcast message \emph{is} received by all nodes within transmission range. This abstraction allows us to interpret a failure of route discovery of a protocol (or any other property) as an imperfection in the protocol, rather than as a result of a chosen formalism not ensuring guaranteed receipt. The same holds for groupcast and unicast messages in case the destinations are within range. (ii) Only internode communication, i.e.\ transferring a message from one node in the network to another, takes time. This is justified as in wireless networks sending a packet takes multiple microseconds; compared to these ``slow'' actions, time spent for internal (intranode) computations, such as variable assignments,  is negligible.
\pagebreak[3]


\section{Modelling OLSRv2 in \tawn}\label{sec:model}
In this section, we present parts of our \tawn model of OLSRv2. The model itself consists of five main processes implementing the OLSRv2 specification and a queue process to receive packets from other routers:

\begin{itemize}[nolistsep]
    \item The process \texttt{OLSR} constitutes the main protocol loop. It is responsible for receiving packets from the input queue and processing these packets according to their type. It it also responsible for periodically generating new HELLO and TC messages.
    \item The \texttt{UPDATE\_INFO} process ensures that the protocol's information bases remain consistent with certain constraints.
    \item The \texttt{PROCESS\_HELLO} and \texttt{PROCESS\_TC} processes are responsible for recording information obtained through HELLO messages and TC messages in the relevant information bases.
    \item The \texttt{FORWARD\_TC} process forwards TC messages received from the router's flooding MPR selectors, subject to a few side conditions.
    \item The \texttt{QUEUE} process receives packets from other routers in the network and delivers them to the \texttt{OLSR} process.
\end{itemize}
Due to a lack of space we only present the process \texttt{OLSR}, including the necessary data structure; the full specification of OLSRv2, including the full data structure and a detailed description can be found in~\autoref{app:OLSR}.

\subsection{Data Structure}
We now describe the data structure needed and functions used in modelling the process \texttt{OLSR}. In the remainder, we use the notation \var{x$_{1..n}$} as shorthand for the tuple \var{(x$_1$,x$_2$,...,x$_n$)}.

We take the type {\tTIME} isomorphic to $\ZZ\cup\{\infty,-\infty\}$. Additionally, we use the type synonym $\texttt{SQN} \mathbin= \typeINT$ for sequence numbers. Moreover, we assume the existence of a polymorphic data type for lists: $$\texttt{data [a] = [] | a:[a]}$$ with the standard functions \texttt{concat} and  \texttt{append} to concatenate two lists and to append a single element to a list, respectively.

The \typeMsg\ data type encompasses both HELLO messages and TC messages, the two communication primitives used by the protocol.
\begin{align*}
    &\texttt{data MESSAGE\ =}\\
    &\texttt{\ \ HELLO \typeIP\ \typeTIME\ \typePOW{\typeIP \times \typeSTATUS}}\ \typePOW{\typeIP \times \typeMPR}\ \typePOW{\typeIP \times \typeMETRIC}\ \typePOW{\typeIP \times \typeMETRIC} \\
    &\texttt{| \texttt{TC \typeIP\ \typeIP\ \typeTIME\ \typeSQN\ \typeSQN\ \typePOW{\typeIP \times \typeMETRIC}}}
\end{align*}
In our implementation, HELLO messages contain six elements
\begin{enumerate*}[label=(\emph{\roman*}), before=\unskip{: }, itemjoin={{; }}, itemjoin*={{, and }}]
    \item an originator address, of the \tawn basic type $\typeIP$
    \item a validity time detailing how long the message's contents should be considered valid for once the message is processed
    \item a set of tuples assigning link statuses to each of the originating router's neighbours
    \item a set of tuples indicating which neighbours the originating router has selected as flooding and routing MPRs
    \item a set of tuples representing the incoming metrics to the originating router from each of its neighbouring routers
    \item a set of tuples representing the outgoing metrics from the originating router to its neighbours.
\end{enumerate*}

TC messages also contain six elements
\begin{enumerate*}[label=(\emph{\roman*}), before=\unskip{: }, itemjoin={{; }}, itemjoin*={{, and }}]
    \item an originator address
    \item a sender address for the router that last broadcast the message
    \item a validity time
    \item a sequence number identifying the message
    \item an advertising neighbour sequence number indicating how fresh the advertised information in the message is
    \item a set of advertised links between the originating router and its routing MPR selectors.
\end{enumerate*}

Routers combine these messages into packets when broadcasting them. We do not care about the packet header, and so take the \typeMSG\ type of \tawn to represent a list of \typeMsg\ values. We determine whether a message is a HELLO message or a TC message by invoking the \var{isHELLO} and \var{isTC} functions.

Routers must be able to generate new messages periodically.
To this end, we use the functions \texttt{newHELLO} and \texttt{newTC}. 
Both make use of a set \texttt{ls} containing link tuples (elements of a data type $\typeL$) that contain information such as an originator address and a validity time; functions of the form \texttt{L\_*} are used to distil information from link tuples. The precise definitions are given in \autoref{app:data}.
\begin{align*}
    &\texttt{newHELLO ::\texttt{\ }$\typeIP \times \typeTIME \times \typePOW{\typeL} \times \typeTIME \rightarrow \typeMsg$} \\[-.2em]
    &\texttt{newHELLO(ip,vtime,ls,now) $\equiv$}\\[-.2em]
    &\LETand \var{statuses} = \setl \var{(L\_oip(lt),L\_status(lt,now))}\ |\ \var{lt} \in \var{ls} \setr \\[-.2em]
    &\AND \var{mprs} = \\[-.2em]
&\hphantom{\AND}
     \setl \var{(L\_oip(lt),FLOODING)}\ |\ \var{lt} \in \var{ls} \wedge \var{L\_fmpr(lt)} \wedge \neg \var{L\_rmpr(lt)} \setr\ \cup \\[-.2em]
&\hphantom{\AND}
    \setl \var{(L\_oip(lt),ROUTING)}\ |\ \var{lt} \in \var{ls} \wedge \neg \var{L\_fmpr(lt)} \wedge \var{L\_rmpr(lt)} \setr\ \cup \\[-.2em]
&\hphantom{\AND}
    \setl \var{(L\_oip(lt),FLOOD\_ROUTE)}\ |\ \var{lt} \in \var{ls} \wedge \var{L\_fmpr(lt)} \wedge \var{L\_rmpr(lt)} \setr \\[-.2em]
    &\AND \var{in\_metrics} = \\[-.2em]
&\hphantom{\AND}
    \setl \var{(L\_oip(lt),L\_in\_metric(lt))}\ |\ \var{lt} \in \var{ls} \wedge \var{L\_status(lt,now)} \neq \var{LOST}  \setr \\[-.2em]
    &\AND \var{out\_metrics} = \\[-.2em]
&\hphantom{\AND}
    \setl \var{(L\_oip(lt),L\_out\_metric(lt))}\ |\ \var{lt} \in \var{ls} \wedge \var{L\_status(lt,now)} = \var{SYMMETRIC} \setr \\[-.2em]
    &\INand \var{HELLO\ ip\ vtime\ statuses\ mprs\ in\_metrics\ out\_metrics}\\[3mm]
    &\texttt{newTC ::\texttt{\ }$\typeIP \times \typeTIME \times \typeSQN \times \typeSQN \times \typePOW{\typeL} \times \typeTIME \rightarrow \typeMsg$} \\
    &\texttt{newTC(ip,vtime,sqn,ansn,ls,now) $\equiv$}\\
    &\LET \var{dests} = \setl \var{(L\_oip(lt),L\_out\_metric(lt))}\ |\ \var{lt} \in \var{ls} \wedge \var{L\_rmpr\_selector(lt)}\ \wedge \\[-.4em]
&\hphantom{\LET \var{dests} = \setl \var{(L\_oip(lt),L\_out\_metric(lt))}\ |\ }
                                                                      \var{L\_status(lt,now) = \var{SYMMETRIC}}\ \setr \\
    &\IN \var{TC\ ip\ ip\ vtime\ sqn\ ansn\ dests}
\end{align*}

\noindent Our protocol maintains its state in a list of variables.
Among others these include
\begin{enumerate}[label=--]
    \item\var{ls}, a link set maintaining information about 1-hop neighbours and their statuses
    \item\var{2hs}, the 2-hop set maintaining information about 2-hop neighbours
    \item\var{arrs}, a remote router set containing information about routers which have advertised links
    \item\var{rts}, the router topology set containing advertised links
    \item\var{rs}, a routing set containing shortest known routes
    \item\var{ps},  the processed set identifying processed TC messages
    \item\var{rxs}, the received set identifying TC messages received and considered for forwarding
    \item\var{pkt}, a list of messages requiring sending, such as those generated by the router or forwarded by it 
    \item\var{hello\_time}, the time when the next HELLO message must be added to \var{pkt}
    \item\var{tc\_time}, the time when the next TC message must be added to \var{pkt}
    \item\var{send\_time}, the time when \var{pkt} must be broadcast
    \item\var{mqueue}, the queue of to-be-processed messages
    \item\var{sqn}, the sequence number identifying a TC message 
    \item\var{ansn}, the advertising neighbour sequence number included in TC messages to indicate how recent the advertised contents are 
    \item\var{prev\_ls}, the previous link set used to check for updates.
\end{enumerate}

\pagebreak

\noindent
These variables are modified during the protocol's execution. For our specification, presented
below,  we use the shortcut $\sigma$ for these 
variables:\vspace{-3mm}
\begin{align*}
\sigma \equiv&\ \var{ls,2hs,arrs,rts,rs,ps,rxs,pkt,hello\_time,tc\_time,send\_time,mqueue,sqn,}\\[-.4em]
&\ \var{ansn,prev\_ls}
\end{align*}
\vspace{-2em}

\noindent
The protocol also maintains variables that are not changed by the protocol itself, including parameters set by a network administrator.  These are 
\begin{enumerate}[label=--]
    \item \var{ip}, the (unique) address of the router
    \item \var{hp\_maxjitter}, the maximum jitter time for HELLO messages
    \item \var{tp\_maxjitter}, the maximum jitter time for TC messages 
    \item \var{h\_hold\_time}, the validity time for generated HELLO messages 
    \item \var{t\_hold\_time}, the validity time for generated TC messages 
    \item \var{l\_hold\_time}, the length that lost links should be kept for 
    \item \var{hello\_interval}, the period between HELLO message transmissions 
    \item \var{tc\_interval}, the period between TC message transmissions,
\pagebreak[3]
\end{enumerate}
and are abbreviated by $\Gamma$.
All variables contained in $\sigma$ and $\Gamma$ are also summarised in \autoref{table:variables}.

Besides these variables, we also maintain a variable \texttt{queue} of type \typeList{\typeMSG}\ in our input queue process and a variable \texttt{msg} of type \typeMsg\ when processing or forwarding a message.

In our model, we localise all relevant information base updates in the single process \texttt{UPDATE\_INFO} (Process \ref{proc:update} in the appendix). We use a condition \var{Updated}, also defined in the appendix, to check whether this process needs to be called. \var{Updated} holds iff the updates of Process \texttt{UPDATE\_INFO} would not modify the protocol's information bases. This is the case iff the information bases are currently consistent, implying for instance that links whose expiration time has elapsed have been purged from the system.

\subsection{The Formal Model}
We present the formal model of the main routine \texttt{OLSR} of OLSRv2 in detail;  it is depicted in \autoref{proc:olsr}.

The main OLSR routine performs a number of different roles. The most basic of these is receiving a packet from the input queue, which occurs in the block on lines \hyperref[proc:olsr]{1-3}. Packets of type \texttt{MSG} are simply lists of HELLO and TC messages, so we concatenate received packets with an existing queue of to-be-processed messages. When the protocol is ready to process a HELLO or TC message, the choice on line~\hyperref[proc:olsr]{8} is taken, with \var{msg} and \var{msgs} assigned to the head and tail of \var{mqueue} respectively by the guard. Note that this guard contains an \texttt{Updated} conjunct, which asserts that the information bases have already been updated by the block on lines \hyperref[proc:olsr]{5-6}. Once \var{msg} is assigned to the element at the head of the queue, the ensuing assignment statement assigns \texttt{mqueue} to its tail \var{msgs}. The guards on lines \hyperref[proc:olsr]{9} and \hyperref[proc:olsr]{12} then ensure that \texttt{msg} is processed according to its type, be that a HELLO message or a TC message.

\begin{algorithm}
{\small
\SetAlgoNoLine
\caption{The main OLSR process\label{proc:olsr}}
\vspace{1pt}
\OLSRDEF{\ensuremath{\sigma,\Gamma}}
\vspace{1pt}
\tcc{Receive a packet (i.e. a list of messages) from the queue process}
\procRECV{\var{msgs}}{
    \procASSN{\var{mqueue} := \var{concat(mqueue,msgs)}}{}
    \procDEFN{OLSR}{\sigma,\Gamma}
}
\vspace{-.15em}
\procCHOICE
\tcc{Execute pending updates to relevant information bases}
\procGUARD{\neg\var{Updated}}{
    \procDEFN{UPDATE\_INFO}{\sigma,\Gamma}\\
}
\vspace{-.15em}
\procCHOICE
\tcc{Process a received message}
\procGGUARD{\var{Updated}\ \wedge\ \var{send\_time} \neq \var{now} \wedge\ \var{mqueue} = (\var{msg}:\var{msgs})}{}
\procASSN{\var{mqueue} := \var{msgs}}{
    \tcc{Process a received HELLO message}
    \procGUARD{\var{isHELLO(msg)}}{
    \procDEFN{PROCESS\_HELLO}{\sigma,\Gamma,\var{msg}}\\
    }
    \vspace{-.15em}
    \procCHOICE
    \tcc{Process a received TC message}
    \procGUARD{\var{isTC(msg)}}{
    \procDEFN{PROCESS\_TC}{\sigma,\Gamma,\var{msg}}\\
    }
}
\vspace{-.15em}
\procCHOICE
\tcc{Time to generate a HELLO message}
\procGUARD{\var{Updated}\ \wedge\ \var{send\_time} \neq \var{now} \wedge\ \var{hello\_time} - \var{hp\_maxjitter} \le \var{now} \le \var{hello\_time}}{
    \tcc{Add the message to the current packet}
    \procASSN{\var{pkt} := \var{append(pkt,newHELLO(ip,h\_hold\_time,ls,now))}}{}
    \tcc{Set relevant timers}
    \procASSN{\var{hello\_time} := \var{now} + \var{hello\_interval}}{}
    \procASSN{\var{send\_time} := \var{now}+1}{}
    \procDEFN{OLSR}{\sigma,\Gamma}
}
\vspace{-.15em}
\procCHOICE
\tcc{Time to generate a TC message}
\procGUARD{\var{Updated}\ \wedge\ \var{send\_time} \neq \var{now} \wedge\ \var{tc\_time} - \var{tp\_maxjitter} \le \texttt{now} \le \var{tc\_time}}{
    \tcc{Add the message to the current packet}
    \procASSN{\var{pkt} := \var{append(pkt,newTC(ip,t\_hold\_time,sqn,ansn,ls,now))}}{}
    \tcc{Increment the sequence number}
    \procASSN{\var{sqn} := \var{sqn} + 1}{}
    \tcc{Set relevant timers}
    \procASSN{\var{tc\_time} := \var{now} + \var{tc\_interval}}{}
    \procASSN{\var{send\_time} := \var{now}+1}{}
    \procDEFN{OLSR}{\sigma,\Gamma}
}
\vspace{-.15em}
\procCHOICE
\tcc{Broadcast the accumulated packet}
\procGUARD{\var{Updated}\ \wedge\ \texttt{send\_time} = \texttt{now}}{
    \procASSN{\texttt{send\_time} := \infty}{}
    \procBC{\var{pkt}}{}
    \procASSN{\var{pkt} := [\hspace{.3em}]}{}
    \procDEFN{OLSR}{\sigma,\Gamma}
}
\vspace{1pt}
}
\end{algorithm}

Aside from processing messages, a router must generate its own HELLO and TC messages periodically. The guard on line \hyperref[proc:olsr]{15} asserts that a new HELLO message is ready to be added to \var{pkt}, a list which accumulates all messages generated during a time tick in order to circumvent broadcasting delays. The guard is true whenever the local clock \texttt{now} enters the jitter period before the message's preparation deadline \texttt{hello\_time}. The condition $\var{now} \le \var{hello\_time}$ is redundant, as one can   prove that it is always met even when left out; it reminds us that we have not yet exceeded the deadline $\var{hello\_time}$ when line   \hyperref[proc:olsr]{16} is executed. A new HELLO message is generated by line \hyperref[proc:olsr]{16} and appended to \texttt{pkt}. Then, \texttt{hello\_time} is increased to \texttt{hello\_interval} time units away from \var{now}, and \texttt{send\_time} is set to $\var{now}+1$ so that the packet will be broadcast during the next tick. Sending a TC message in the block starting at line \hyperref[proc:olsr]{21} follows a similar process, except that a sequence number \var{sqn} is included in the message and subsequently incremented. Once all messages have been accumulated and the guard on line \hyperref[proc:olsr]{28} becomes true, the accumulated packet is broadcast, and both \texttt{pkt} and \texttt{send\_time} are reset to indicate no pending messages.

\section{Correcting the Specification}
\label{sec:bug}
As currently specified, the OLSRv2 model does not guarantee that optimal routes will be established to all reachable destinations. 
This is in contrast to the intention of the RFC. The reason is that incorrect directional link metrics are being used in
section 18.5 of RFC 7181 \cite{rfc7181} during routing MPR selection. Rather than using the incoming metric between 1-hop and 2-hop neighbours, the outgoing metric from the 1-hop neighbour to the 2-hop neighbour is used instead:
\begin{quote}\renewcommand*\ttdefault{lmvtt}
\begin{ttfamily}{\rm ``}For each element x in N1, define N2(x) as the set of elements y in 
         N2 whose corresponding address is the N2\_2hop\_addr of an allowed 
        2-Hop Tuple that has N2\_neighbor\_iface\_addr\_list contained in  
         N\_neighbor\_addr\_list of the Neighbor Tuple corresponding to x. For
         all such x and y, define d2(x,y) := {\color{red}{N2\_out\_metric}} of that 2-Hop 
         Tuple.{\rm''}
\end{ttfamily}
\end{quote}

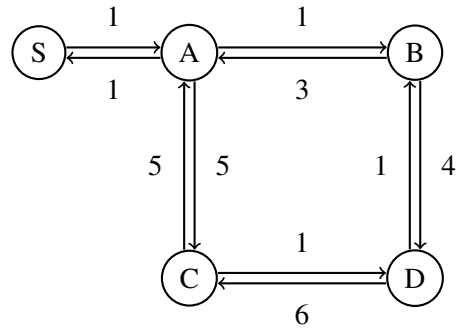
\begin{figure}
\centering
\begin{tikzpicture}
    \node[shape=circle,thick,draw=black] (S) at (-2,3) [] {S};
    \node[shape=circle,thick,draw=black] (B) at (3,3) [] {B};
    \node[shape=circle,thick,draw=black] (A) at (0,3) [] {A};
    \node[shape=circle,thick,draw=black] (D) at (3,0) [] {D};
    \node[shape=circle,thick,draw=black] (C) at (0,0) [] {C};
    \path [->,thick,transform canvas={yshift=-.4ex}] (A) edge node[label=below:{1}] {} (S);
    \path [->,thick,transform canvas={yshift=.4ex}] (S) edge node[label=above:{1}] {} (A);
    \path [->,thick,transform canvas={yshift=-.4ex}] (B) edge node[label=below:{3}] {} (A);
    \path [->,thick,transform canvas={yshift=.4ex}] (A) edge node[label=above:{1}] {} (B);
    \path [->,thick,transform canvas={yshift=-.4ex}] (D) edge node[label=below:{6}] {} (C);
    \path [->,thick,transform canvas={yshift=.4ex}] (C) edge node[label=above:{1}] {} (D);
    \path [->,thick,transform canvas={xshift=-.4ex}] (C) edge node[label=left:{5}] {} (A);
    \path [->,thick,transform canvas={xshift=.4ex}] (A) edge node[label=right:{5}] {} (C);
    \path [->,thick,transform canvas={xshift=-.4ex}] (D) edge node[label=left:{1}] {} (B);
    \path [->,thick,transform canvas={xshift=.4ex}] (B) edge node[label=right:{4}] {} (D);
\end{tikzpicture}
\caption{A counterexample to route optimality}
\label{fig:counter}
\end{figure}

From this definition, we can construct a straightforward counterexample. 
Consider the topology in \autoref{fig:counter} and assume that each node is aware of the links to its 1-hop neighbours and 2-hop neighbours. To ensure that S has a shortest path to D, D must ensure that its sole 2-hop neighbour, A, has a shortest path to D. When selecting a routing MPR to advertise an appropriate link, D compares the cost of the paths from A via C and from A via B. However, it uses the outgoing metric when calculating the cost of a link from A to C and from A to B. The path via C, which has a cost of $5 + 1 = 6$, is ``shorter'' than the path via B, which has a cost of $3 + 4 = 7$. Therefore, D will require the path from A to take the ``shorter'' route via C. When D sends its next HELLO message, it tells C to advertise this link, which C subsequently includes in its TC messages. Eventually, S learns of the links from A to C, A to B and C to D via TC messages. However, it never learns of the link from B to D since D does not tell B to advertise on its behalf. Therefore, S will not be able to construct the true shortest path to D via B.

This error is straightforward to correct, simply by replacing \texttt{N2\_out\_metric} with \texttt{N2\_in\_metric}. The ``bug'' is already corrected in the model presented in the appendix.

Note that this bug does not affect the \url{OLSR.org} open source implementation of OLSR version 2. The OLSR daemon version 2 \cite{OONF}, included in the OLSR.org Network Framework, correctly uses the incoming metric.


\section{Related Work, Discussion \& Future Work}

This paper provides a full and detailed model of the routing protocol OLSRv2, written in the process algebra \tawn~\cite{ESOP16}. Thanks to the formal semantics of \tawn, this model is completely unambiguous and forms a good basis for the verification of useful protocol properties. Moreover, it is considerably shorter than the $203$ pages of English prose present in the RFCs for OLSRv2~\cite{rfc6130} and its base  NHDP~\cite{rfc7181}.
As with all formal models, our model represents our interpretation of the English prose; 
a formal proof that a model or an implementation is compliant with a textual specification is impossible.
However, the existence of a detailed, unambiguous model, such as ours, can be the base for a formal refinement proof 
that an implementation is compliant to that model. The fact that our model is written in ``pseudo-code'' would make such a 
proof easier compared to other modelling languages.

To the best of our knowledge, our model is by far the most detailed model of the Optimised Link State Routing protocol found in the literature---we believe that it is the very first model of version 2.  There have only been a few other efforts to formally model OLSR. 

Baras et al.\ \cite{BTPS09} present a component-based methodology for modelling and designing
wireless routing protocols, and illustrate it on OLSR\@. Although many crucial aspects of the
protocol, such as MPR selection, are captured this way, the result cannot be seen as a full
rendering of the core functionality of the protocol. Consequently, it could not be used as basis
for the verification of properties like route optimality.

Steele and Andel \cite{SA12} provide a model of OLSR for analysis with the model checker SPIN.
The model abstracts, among others, from timing aspects, and thus cannot detect possible shortcomings
of the protocol resulting, for example, from premature route expiry.

Kamali et al.\ \cite{KHKP15} model OLSR in the input language of the model checker {\sc Uppaal}.
To facilitate model checking, the model abstracts from large parts of the information bases of OLSR\@.
To avoid the computation of shortest paths they update the routing table whenever a TC message is received.
Although this reduces the state space drastically, it changes the protocol behaviour in such a way that 
optimal routes cannot be guaranteed. This problem is a shortcoming of the abstract model rather than the protocol itself.

In \cite{KP16}, Kamali and Petre model OLSR in the state-based formal method \emph{Event-B}.
This model abstracts from timing, but is otherwise largely consistent with the above {\sc Uppaal} model.

Currently, there is a limited amount of tool support available for \tawnopt. 
The untimed fragment has been integrated into the interactive proof assistant Isabelle/HOL~\cite{BGH16,AWN-AFP}.
Specifications formalised in \tawnopt can also be a base for model checking: in~\cite{GHW18} it is shown how to 
analyse AWN-specifications using \emph{mCRL2}; a translation from \tawnopt into {\sc Uppaal} is sketched in~\cite{TACAS12} and 
illustrated in \cite{DHW20} for the routing protocol OSPF.

The one feature our model of OLSRv2 abstracts from is that routers may have multiple communication interfaces. Adding support for multiple interfaces is a possible topic for future work; it might, however, necessitate extending the expressiveness of \tawn.

Future work additionally involves employing the model provided here for the verification of requirements a routing protocol like OLSRv2 should satisfy: \emph{route correctness}, saying that all routes found by the protocol actually exist, \emph{route discovery}, saying that when a route between two nodes exists, a route will be found by the protocol, and \emph{route optimality}, saying that the protocol finds the best routes possible.

\newpage

\providecommand{\thisvolume}[2][ggg]{this volume of EPTCS, Open Publishing Association}
\bibliographystyle{eptcs}
\bibliography{olsr}
\newpage
\appendix

\section{Full and Detailed Model of OLSRv2\label{app:OLSR}}
In this appendix we provide a full specification of the routing protocol OLSRv2~\cite{rfc7181}, 
written in the formal language \tawn~\cite{ESOP16}. In \autoref{app:data} we present the full data structure (types, variables, functions, etc.) needed to formally specify the protocol in \autoref{app:model}.


\setcounter{table}{2}

\subsection{Data Structure\label{app:data}}
In this section, we describe the data types and functions used in modelling OLSRv2.
We define some basic types in addition to those required by \tawn, provide functions that operate on these types, and finally list the variables which form the protocol's state.

In the remainder, we use the notation \var{x$_{1..n}$} as shorthand for the tuple \var{(x$_1$,x$_2$,...,x$_n$)}.
Moreover, we use $\dexp{e} \updl \var{x} := \dexp{t} \updr$ for syntactic substitution.
Here, all instances of \var{x}\ in $\dexp{e}$ are replaced with $\dexp{t}$.

\subsubsection*{Basic Types}

We begin by defining a few basic data types and type synonyms.
\begin{align*}
    &\texttt{data STATUS = SYMMETRIC | HEARD | LOST}\\
    &\texttt{data MPR = FLOODING | ROUTING | FLOOD\_ROUTE}\\
    &\texttt{type SQN = \typeINT}\\
    &\texttt{type METRIC = $\typeNAT_{>0} \cup \setl \infty \setr$}
\end{align*}
Moreover, we assume the existence of a polymorphic data type for lists.
\[
\texttt{data [a] = [] | a:[a]}
\]
\noindent We enable the concatenation of two lists and appending to a list via the functions
 \texttt{concat},
and  \texttt{append},
respectively.

\subsubsection*{Messages}

The \typeMsg\ data type encompasses both HELLO messages and TC messages, the two communication primitives used by the protocol.
\begin{align*}
    &\texttt{data MESSAGE\ =}\\
    &\texttt{\ \ HELLO \typeIP\ \typeTIME\ \typePOW{\typeIP \times \typeSTATUS}}\ \typePOW{\typeIP \times \typeMPR}\ \typePOW{\typeIP \times \typeMETRIC}\ \typePOW{\typeIP \times \typeMETRIC} \\
    &\texttt{| \texttt{TC \typeIP\ \typeIP\ \typeTIME\ \typeSQN\ \typeSQN\ \typePOW{\typeIP \times \typeMETRIC}}}
\end{align*}
In our implementation, HELLO messages contain six elements
\begin{enumerate*}[label=(\emph{\roman*}), before=\unskip{: }, itemjoin={{; }}, itemjoin*={{, and }}]
    \item an originator address uniquely identifying the router that generated the message
    \item a validity time detailing how long the message's contents should be considered valid for once the message is processed
    \item a set of tuples assigning link statuses to each of the originating router's neighbours
    \item a set of tuples indicating which neighbours the originating router has selected as flooding and routing MPRs
    \item a set of tuples representing the incoming metrics to the originating router from each of its neighbouring routers
    \item a set of tuples representing the outgoing metrics from the originating router to its neighbours.
\end{enumerate*}

TC messages also contain six elements
\begin{enumerate*}[label=(\emph{\roman*}), before=\unskip{: }, itemjoin={{; }}, itemjoin*={{, and }}]
    \item an originator address uniquely identifying the router that generated the message
    \item a sender address uniquely identifying the router that last broadcast the message
    \item a validity time detailing how long the message's contents should be considered valid for once the message is processed
    \item a sequence number which, when used in tandem with the originator address, uniquely identifies the message
    \item an advertising neighbour sequence number indicating how fresh the advertised information in the message is
    \item a set of advertised links between the originating router and its routing MPR selectors.
\end{enumerate*}

Routers combine these messages into packets when broadcasting them.
We do not care about the packet header, and so take the \typeMSG\ type of T-AWN to represent a list of \typeMsg\ values.
$$\texttt{type MSG = [MESSAGE]}$$
We determine whether a message is a HELLO message or a TC message by invoking
the \var{isHELLO} and \var{isTC} functions.
\begin{align*}
    &\texttt{isHELLO ::\texttt{\ }$\typeMsg \rightarrow \typeBOOL$} \\
    &\texttt{isHELLO(HELLO \_ \_ \_ \_ \_ \_)\ $\equiv$ \TRUE}\\
    &\texttt{isHELLO(TC \_ \_ \_ \_ \_ \_)\ \ \ \ $\equiv$ \FALSE}
    \\[3mm]
    &\texttt{isTC ::\texttt{\ }$\typeMsg \rightarrow \typeBOOL$} \\
    &\texttt{isTC(HELLO \_ \_ \_ \_ \_ \_)\ $\equiv$ \FALSE}\\
    &\texttt{isTC(TC \_ \_ \_ \_ \_ \_)\ \ \ \ $\equiv$ \TRUE}
\end{align*}
We also use functions to extract the contents of these messages.
Both HELLO and TC messages contain an originator address and validity time,
which we access via the \var{oip} and \var{vtime} functions.
\begin{align*}
    &\texttt{oip ::\texttt{\ }$\typeMsg \rightarrow \typeIP$} \\
    &\texttt{oip(HELLO x$_1$ \_ \_ \_ \_ \_)\ $\equiv$ x$_1$}\\
    &\texttt{oip(TC x$_1$ \_ \_ \_ \_ \_)\hphantom{LLO\ }$\equiv$ x$_1$}
    \\[3mm]
    &\texttt{vtime ::\texttt{\ }$\typeMsg \rightarrow \typeTIME$} \\
    &\texttt{vtime(HELLO \_ x$_2$ \_ \_ \_ \_)\ $\equiv$ x$_2$}\\
    &\texttt{vtime(TC \_ \_ x$_3$ \_ \_ \_)\ \ \ \ $\equiv$ x$_3$}
\end{align*}
HELLO messages contain some unique elements that are not present in TC messages.
We define a number of partial functions for extracting these elements.
\begin{align*}
    &\texttt{statuses ::\texttt{\ }$\typeMsg \rightharpoonup \typePOW{\typeIP \times \typeSTATUS}$ } \\
    &\texttt{statuses(HELLO \_ \_ x$_3$ \_ \_ \_)\ $\equiv$ x$_3$}
    \\[3mm]
    &\texttt{mprs ::\texttt{\ }$\typeMsg \rightharpoonup \typePOW{\typeIP \times \typeMPR}$ } \\
    &\texttt{mprs(HELLO \_ \_ \_ x$_4$ \_ \_)\ $\equiv$ x$_4$}
    \\[3mm]
    &\texttt{inMetrics ::\texttt{\ }$\typeMsg \rightharpoonup \typePOW{\typeIP \times \typeMETRIC}$ } \\
    &\texttt{inMetrics(HELLO \_ \_ \_ \_ x$_5$ \_)\ $\equiv$ x$_5$}
\end{align*}

\pagebreak

\noindent\begin{align*}
    &\texttt{outMetrics ::\texttt{\ }$\typeMsg \rightharpoonup \typePOW{\typeIP \times \typeMETRIC}$ } \\
    &\texttt{outMetrics(HELLO \_ \_ \_ \_ \_ x$_6$)\ $\equiv$ x$_6$}
\end{align*}
Similarly, we define some partial functions for extracting the unique elements of TC messages.
\begin{align*}
    &\texttt{sip ::\texttt{\ }$\typeMsg \rightharpoonup \typeIP$ } \\
    &\texttt{sip(TC \_ x$_2$ \_ \_ \_ \_)\ $\equiv$ x$_2$}
    \\[3mm]
    &\texttt{sqn ::\texttt{\ }$\typeMsg \rightharpoonup \typeSQN$ } \\
    &\texttt{sqn(TC \_ \_ \_ x$_4$ \_ \_) $\equiv$ x$_4$}
    \\[3mm]
    &\texttt{ansn ::\texttt{\ }$\typeMsg \rightharpoonup \typeSQN$ } \\
    &\texttt{ansn(TC \_ \_ \_ \_ x$_5$ \_)\ $\equiv$ x$_5$}
    \\[3mm]
    &\texttt{dests ::\texttt{\ }$\typeMsg \rightharpoonup \typePOW{\typeIP \times \typeMETRIC}$ } \\
    &\texttt{dests(TC \_ \_ \_ \_ \_ x$_6$)\ $\equiv$ x$_6$}
\end{align*}
 
In addition to manipulating existing messages, routers must be able to generate new messages periodically.
To this end, the functions \texttt{newHELLO} and \texttt{newTC} are used to generate new HELLO and TC messages respectively.
Both make use of a set \texttt{ls} containing values of type $\typeL$ (defined on \autopageref{sec:iib}).
\begin{align*}
    &\texttt{newHELLO ::\texttt{\ }$\typeIP \times \typeTIME \times \typePOW{\typeL} \times \typeTIME \rightarrow \typeMsg$} \\[-.2em]
    &\texttt{newHELLO(ip,vtime,ls,now) $\equiv$}\\[-.2em]
    &\LETand \var{statuses} = \setl \var{(L\_oip(lt),L\_status(lt,now))}\ |\ \var{lt} \in \var{ls} \setr \\[-.2em]
    &\AND \var{mprs} = \\[-.2em]
&\hphantom{\AND}
     \setl \var{(L\_oip(lt),FLOODING)}\ |\ \var{lt} \in \var{ls} \wedge \var{L\_fmpr(lt)} \wedge \neg \var{L\_rmpr(lt)} \setr\ \cup \\[-.2em]
&\hphantom{\AND}
    \setl \var{(L\_oip(lt),ROUTING)}\ |\ \var{lt} \in \var{ls} \wedge \neg \var{L\_fmpr(lt)} \wedge \var{L\_rmpr(lt)} \setr\ \cup \\[-.2em]
&\hphantom{\AND}
    \setl \var{(L\_oip(lt),FLOOD\_ROUTE)}\ |\ \var{lt} \in \var{ls} \wedge \var{L\_fmpr(lt)} \wedge \var{L\_rmpr(lt)} \setr \\[-.2em]
    &\AND \var{in\_metrics} = \\[-.2em]
&\hphantom{\AND}
    \setl \var{(L\_oip(lt),L\_in\_metric(lt))}\ |\ \var{lt} \in \var{ls} \wedge \var{L\_status(lt,now)} \neq \var{LOST}  \setr \\[-.2em]
    &\AND \var{out\_metrics} = \\[-.2em]
&\hphantom{\AND}
    \setl \var{(L\_oip(lt),L\_out\_metric(lt))}\ |\ \var{lt} \in \var{ls} \wedge \var{L\_status(lt,now)} = \var{SYMMETRIC} \setr \\[-.2em]
    &\INand \var{HELLO\ ip\ vtime\ statuses\ mprs\ in\_metrics\ out\_metrics}
    \\[3mm]
    &\texttt{newTC ::\texttt{\ }$\typeIP \times \typeTIME \times \typeSQN \times \typeSQN \times \typePOW{\typeL} \times \typeTIME \rightarrow \typeMsg$} \\
    &\texttt{newTC(ip,vtime,sqn,ansn,ls,now) $\equiv$}\\
    &\LET \var{dests} = \setl \var{(L\_oip(lt),L\_out\_metric(lt))}\ |\ \var{lt} \in \var{ls} \wedge \var{L\_rmpr\_selector(lt)}\ \wedge \\[-.4em]
&\hphantom{\LET \var{dests} = \setl \var{(L\_oip(lt),L\_out\_metric(lt))}\ |\ }
                                                                      \var{L\_status(lt,now) = \var{SYMMETRIC}}\ \setr \\
    &\IN \var{TC\ ip\ ip\ vtime\ sqn\ ansn\ dests}
\end{align*}

Outside of periodic generation,
a TC message may also be forwarded by a router if received from one of that router's flooding MPR selectors.
In this scenario, we use the function \texttt{forward} to update the sender IP of the TC message
to the router's own address before broadcasting it.
\begin{align*}
    &\texttt{forward ::\texttt{\ }$\typeIP \times \typeMSG \rightharpoonup \typeMSG$}\\
    &\texttt{forward(ip,\hspace{.1em}TC oip sip vtime sqn ansn dests)} \equiv\\
    &\texttt{\ \ TC\ oip\ ip\ vtime\ sqn\ ansn\ dests}
\end{align*}

\subsubsection*{Interface Information Base\label{sec:iib}}

The \emph{interface information base} records information about the links between a router and its 1-hop/2-hop neighbours.
\vspace{-1.5ex}
\begin{align*}
    &\texttt{type L\ \ = $\typeIP \times \typeTIME \times \typeTIME \times \typeTIME \times \typeBOOL \times \typeBOOL \times \typeBOOL \times \typeBOOL \times \typeMETRIC \times \typeMETRIC$}\\
    &\texttt{type N2 = $\typeIP \times \typeIP \times \typeTIME \times \typeMETRIC \times \typeMETRIC$}
\end{align*}
A link tuple of type \texttt{L} contained within a router's link set consists of
\begin{enumerate*}[label=(\emph{\roman*}), before=\unskip{: }, itemjoin={{; }}, itemjoin*={{, and }}]
    \item an originator address uniquely identifying the neighbouring router
    \item a symmetric time until which the neighbour should be considered symmetric
    \item a heard time until which the neighbour should be considered heard
    \item a validity time after which the tuple should be removed
    \item a Boolean denoting whether the neighbour is a flooding MPR
    \item a Boolean denoting whether the neighbour is a routing MPR
    \item a Boolean denoting whether the neighbour is a flooding MPR selector
    \item a Boolean denoting whether the neighbour is a routing MPR selector
    \item an incoming link metric from the neighbour to this router
    \item an outgoing link metric to the neighbour from this router.
\end{enumerate*}

A 2-hop tuple of type \texttt{N2} contained within a router's 2-hop set consists of
\begin{enumerate*}[label=(\emph{\roman*}), before=\unskip{: }, itemjoin={{; }}, itemjoin*={{, and }}]
    \item an originator address uniquely identifying the 1-hop neighbour
    \item an originator address uniquely identifying the 2-hop neighbour
    \item a validity time after which the tuple should be removed
    \item an incoming link metric from the 2-hop neighbour to the 1-hop neighbour
    \item an outgoing link metric from the 1-hop neighbour to the 2-hop neighbour.
\end{enumerate*}

Elements of the link set and 2-hop set are accessed through functions of the form \var{L\_*} and \var{N2\_*}.
{\setlength{\tabcolsep}{2em}
\begin{longtable}{l l}
$\begin{aligned}
    &\texttt{L\_oip ::\texttt{\ }$\typeL \rightarrow \typeIP$}\\
    &\texttt{L\_oip(lt$_{1..10}$)} \equiv \var{lt_1}
\end{aligned}$
&
$\begin{aligned}
    &\texttt{L\_symmetric\_time ::\texttt{\ }$\typeL \rightarrow \typeTIME$}\\
    &\texttt{L\_symmetric\_time(lt$_{1..10}$)} \equiv \var{lt_2}
\end{aligned}$
\\[1.5em]
$\begin{aligned}
    &\texttt{L\_heard\_time ::\texttt{\ }$\typeL \rightarrow \typeTIME$}\\
    &\texttt{L\_heard\_time(lt$_{1..10}$)} \equiv \var{lt_3}
\end{aligned}$
&
$\begin{aligned}
    &\texttt{L\_time ::\texttt{\ }$\typeL \rightarrow \typeTIME$}\\
    &\texttt{L\_time(lt$_{1..10}$)} \equiv \var{lt_4}
\end{aligned}$
\\[1.5em]
$\begin{aligned}
    &\texttt{L\_fmpr ::\texttt{\ }$\typeL \rightarrow \typeBOOL$}\\
    &\texttt{L\_fmpr(lt$_{1..10}$)} \equiv \var{lt_5}
\end{aligned}$
&
$\begin{aligned}
    &\texttt{L\_rmpr ::\texttt{\ }$\typeL \rightarrow \typeBOOL$}\\
    &\texttt{L\_rmpr(lt$_{1..10}$)} \equiv \var{lt_6}
\end{aligned}$
\\[1.5em]
$\begin{aligned}
    &\texttt{L\_fmpr\_selector ::\texttt{\ }$\typeL \rightarrow \typeBOOL$}\\
    &\texttt{L\_fmpr\_selector(lt$_{1..10}$)} \equiv \var{lt_7}
\end{aligned}$
&
$\begin{aligned}
    &\texttt{L\_rmpr\_selector ::\texttt{\ }$\typeL \rightarrow \typeBOOL$}\\
    &\texttt{L\_rmpr\_selector(lt$_{1..10}$)} \equiv \var{lt_8}
\end{aligned}$
\\[1.5em]
$\begin{aligned}
    &\texttt{L\_in\_metric ::\texttt{\ }$\typeL \rightarrow \typeMETRIC$}\\
    &\texttt{L\_in\_metric(lt$_{1..10}$)} \equiv \var{lt_9}
\end{aligned}$
&
$\begin{aligned}
    &\texttt{L\_out\_metric ::\texttt{\ }$\typeL \rightarrow \typeMETRIC$}\\
    &\texttt{L\_out\_metric(lt$_{1..10}$)} \equiv \var{lt_{10}}
\end{aligned}$
\end{longtable}}
\vspace{-1em}

{\setlength{\tabcolsep}{1.5em}
\begin{longtable}{l l l}
$\begin{aligned}
    &\texttt{N2\_1h\_oip ::\texttt{\ }$\typeNT \rightarrow \typeIP$}\\
    &\texttt{N2\_1h\_oip(n2$_{1..5}$)} \equiv \var{n2_1}
\end{aligned}$
&
$\begin{aligned}
    &\texttt{N2\_2h\_oip ::\texttt{\ }$\typeNT \rightarrow \typeIP$}\\
    &\texttt{N2\_2h\_oip(n2$_{1..5}$)} \equiv \var{n2_2}
\end{aligned}$
&
$\begin{aligned}
    &\texttt{N2\_time ::\texttt{\ }$\typeNT \rightarrow \typeTIME$}\\
    &\texttt{N2\_time(n2$_{1..5}$)} \equiv \var{n2_3}
\end{aligned}$
\end{longtable}}
\vspace{-1.5em}

{\setlength{\tabcolsep}{1.5em}
\begin{longtable}{l l}
$\begin{aligned}
    &\texttt{N2\_in\_metric ::\texttt{\ }$\typeNT \rightarrow \typeMETRIC$}\\
    &\texttt{N2\_in\_metric(n2$_{1..5}$)} \equiv \var{n2_4}
\end{aligned}$
&
$\begin{aligned}
    &\texttt{N2\_out\_metric ::\texttt{\ }$\typeNT \rightarrow \typeMETRIC$}\\
    &\texttt{N2\_out\_metric(n2$_{1..5}$)} \equiv \var{n2_5}
\end{aligned}$
\end{longtable}}

The status of link tuples is determined as per section 7.1 of RFC 6130 \cite{rfc6130}, although we do not make use of link hysteresis.
If the symmetric time of the tuple is greater than the current time, then the link is considered symmetric.
Else, if the heard time of the tuple is greater than the current time, then the link is considered heard.
Otherwise, the link is considered lost.

\noindent\begin{align*}
    &\texttt{L\_status ::\texttt{\ }$\typeL \times \typeTIME \rightarrow \typeSTATUS$}\\
    &\texttt{L\_status(lt$_{1..10}$,now)} \equiv \\
    &\IF \var{lt_2} > \var{now} \textbf{\ \ \bf then\ \ } \var{SYMMETRIC}\\
    &\ELIF \var{lt_3} > \var{now} \textbf{\ \ \bf then\ \ } \var{HEARD}\\
    &\ELSE \var{LOST}
\end{align*}

When a new HELLO message is received, both the link set and 2-hop set are updated.
The functions \texttt{addLinkTuple} and \texttt{add2HopTuples} create new link tuples and 2-hop tuples if they do not already exist.
In the latter case, there is an additional check to ensure that a symmetric link tuple for the neighbour exists before considering 2-hop neighbours.
We also include functions to update the metrics, timing values and MPR selection statuses of each of the tuples, existing or newly added, based on the message contents.
\begin{align*}
    &\texttt{addLinkTuple ::\texttt{\ }$\typePOW{\typeL} \times \typeIP \times \typeTIME \times \typeMETRIC \times \typeTIME \rightarrow \typePOW{\typeL}$}\\[-.1em]
    &\texttt{addLinkTuple(ls,moip,vtime,in\_metric,now)} \equiv\\[-.1em]
    &\IF \forall \var{lt} \in \var{ls}.\ \var{L\_oip(lt)} \neq \var{moip}\\[-.1em]
    &\THEN \var{ls} \cup \var{\{(moip,-\infty,-\infty,now+vtime,\FALSE,\FALSE,\FALSE,\FALSE,in\_metric,\infty)\}}\\[-.1em]
    &\ELSE \var{ls}
    \\[3mm]
          &\texttt{updateLinkOutMetrics ::\texttt{\ }$\typeIP \times \typePOW{\typeL} \times \typeIP \times \typePOW{\typeIP \times \typeMETRIC} \rightarrow \typePOW{\typeL}$}\\[-.1em]
          &\texttt{updateLinkOutMetrics(ip,ls,moip,in\_metrics)} \equiv\\[-.1em]
          &\texttt{\ \ } \setl \var{lt} \in \var{ls}\ |\ \var{L\_oip(lt)} \neq \var{moip} \vee \forall \var{m_{1..2}} \in \var{in\_metrics}.\ \var{m_1} \neq \var{ip} \setr\ \cup \\[-.1em]
          &\texttt{\ \ } \setl \var{lt_{1..10}} \updl \var{lt_{10}} := \var{m_2} \updr\ |\ \var{lt_{1..10}} \in \var{ls} \wedge \var{L\_oip(lt_{1..10})} = \var{moip}\ \wedge
                               \var{m_{1..2}} \in \var{in\_metrics} \wedge \var{m_1} = \var{ip} \setr
    \\[3mm]
          &\texttt{updateSymmetricTime ::\texttt{\ }$\typeIP \times \typePOW{\typeL} \times \typeIP \times \typeTIME \times \typePOW{\typeIP \times \typeSTATUS} \times \typeTIME \times \typeTIME \rightarrow \typePOW{\typeL}$}\\[-.1em]
          &\texttt{updateSymmetricTime(ip,ls,moip,vtime,statuses,htime,now)} \equiv \\[-.1em]
          &\IF \exists \var{x_{1..2}} \in \var{statuses}.\ \var{x_1} = \var{ip} \wedge
                                \var{x_2} \neq \var{LOST} \\[-.1em]
          &\THEN \setl \var{lt} \in \var{ls}\ |\ \var{L\_oip(lt)} \neq \var{moip} \setr\ \cup\\[-.1em]
&\hphantom{\THEN}
                 \setl \var{lt_{1..10}} \updl \var{lt_2} := \var{now+vtime} \updr\ |\ \var{lt_{1..10}} \in \var{ls} \wedge \var{L\_oip(lt_{1..10}}\var{)} = \var{moip} \setr \\[-.1em]
          &\ELIF \exists \var{x_{1..2}} \in \var{statuses}.\ \var{x_1} = \var{ip} \wedge \var{x_2} = \var{LOST}\ \wedge\\[-.1em]
&\hphantom{\ELIF}
                              \exists \var{lt} \in \var{ls}.\ \var{L\_oip(lt)} = \var{moip} \wedge \var{L\_status(lt,now)} = \var{SYMMETRIC}\\[-.1em]
          &\THEN \setl \var{lt} \in \var{ls}\ |\ \var{L\_oip(lt)} \neq \var{moip} \setr\ \cup\\[-.1em]
&\hphantom{\THEN}
                 \setl \var{lt_{1..10}} \updl \var{lt_2} := -\infty, \var{lt_4} := \var{now+htime} \updr\ |\ \var{lt_{1..10}} \in \var{ls} \wedge \var{L\_oip(lt_{1..10})} = \var{moip} \setr \\[-.1em]
    &\ELSE \var{ls}
    \\[3mm]
    &\texttt{updateHeardTime ::\texttt{\ }$\typePOW{\typeL} \times \typeIP \times \typeTIME \times \typeTIME \rightarrow \typePOW{\typeL}$}\\
    &\texttt{updateHeardTime(ls,moip,vtime,now)} \equiv\\
    &\texttt{\ \ } \setl \var{lt} \in \var{ls}\ |\ \var{L\_oip(lt)} \neq \var{moip} \setr\ \cup\\
    &\texttt{\ \ } \setl \var{lt_{1..10}} \updl \var{lt_3} := \var{max(now+vtime,lt_2)} \updr\ |\ \var{lt_{1..10}} \in \var{ls} \wedge \var{L\_oip(lt_{1..10})} = \var{moip} \setr
    \\[3mm]
    &\texttt{updateValidityTime ::\texttt{\ }$\typePOW{\typeL} \times \typeIP \times \typeTIME \times \typeTIME \rightarrow \typePOW{\typeL}$}\\
    &\texttt{updateValidityTime(ls,moip,htime,now)} \equiv\\
    &\texttt{\ \ } \setl \var{lt} \in \var{ls}\ |\ \var{L\_oip(lt)} \neq \var{moip} \setr\ \cup\\
    &\texttt{\ \ } \setl \var{lt_{1..10}} \updl \var{lt_4} := \var{max(lt_3+htime,lt_4)} \updr \ |\ \var{lt_{1..10}} \in \var{ls} \wedge \var{L\_oip(lt_{1..10})} = \var{moip} \setr
\end{align*}
\begin{align*}
          &\texttt{updateFMPRSelectors ::\texttt{\ }$\typeIP \times \typePOW{\typeL} \times \typeIP \times \typePOW{\typeIP \times \typeMETRIC} \times \typePOW{\typeIP \times \typeMPR} \times \typeTIME \rightarrow \typePOW{\typeL}$}\\
          &\texttt{updateFMPRSelectors(ip,ls,moip,statuses,mprs,now)} \equiv \\
          &\IF \exists \var{x_{1..2}} \in \var{mprs}.\ \var{x_1} = \var{ip} \wedge \var{x_2} \in \setl \var{FLOODING},\var{FLOOD\_ROUTE} \setr\\
          &\THEN \setl \var{lt} \in \var{ls}\ |\ \var{L\_oip(lt)} \neq \var{moip} \setr\ \cup \\[-.2em]
          &\hphantom{\THEN}
                \setl \var{lt_{1..10}} \updl \var{lt_7} := \TRUE \updr\ |\ \var{lt_{1..10}} \in \var{ls} \wedge \var{L\_oip(lt_{1..10})} = \var{moip} \setr\\
          &\ELIF \exists \var{x_{1..2}} \in \var{statuses}.\ \var{x_1} = \var{ip} \wedge \var{x_2} = \var{SYMMETRIC}\\
          &\THEN \setl \var{lt} \in \var{ls}\ |\ \var{L\_oip(lt)} \neq \var{moip} \setr\ \cup \\[-.2em]
          &\hphantom{\THEN}
                 \setl \var{lt_{1..10}} \updl \var{lt_7} := \FALSE \updr \ |\ \var{lt_{1..10}} \in \var{ls} \wedge \var{L\_oip(lt_{1..10})} = \var{moip} \setr \\
          &\ELSE \var{ls}
    \\[3mm]
          &\texttt{updateRMPRSelectors ::\texttt{\ }$\typeIP \times \typePOW{\typeL} \times \typeIP \times \typePOW{\typeIP \times \typeMETRIC} \times \typePOW{\typeIP \times \typeMPR} \times \typeTIME \rightarrow \typePOW{\typeL}$}\\
          &\texttt{updateRMPRSelectors(ip,ls,moip,statuses,mprs,now)} \equiv \\
          &\IF \exists \var{x_{1..2}} \in \var{mprs}.\ \var{x_1} = \var{ip} \wedge \var{x_2} \in \setl \var{ROUTING},\var{FLOOD\_ROUTE} \setr\\
          &\THEN \setl \var{lt} \in \var{ls}\ |\ \var{L\_oip(lt)} \neq \var{moip} \setr\ \cup \\[-.2em]
          &\hphantom{\THEN}
                \setl \var{lt_{1..10}} \updl \var{lt_8} := \TRUE \updr\ |\ \var{lt_{1..10}} \in \var{ls} \wedge \var{L\_oip(lt_{1..10})} = \var{moip} \setr\\
          &\ELIF \exists \var{x_{1..2}} \in \var{statuses}.\ \var{x_1} = \var{ip} \wedge \var{x_2} = \var{SYMMETRIC}\\
          &\THEN \setl \var{lt} \in \var{ls}\ |\ \var{L\_oip(lt)} \neq \var{moip} \setr\ \cup \\[-.2em]
          &\hphantom{\THEN}
                \setl \var{lt_{1..10}} \updl \var{lt_8} := \FALSE \updr \ |\ \var{lt_{1..10}} \in \var{ls} \wedge \var{L\_oip(lt_{1..10})} = \var{moip} \setr \\
          &\ELSE \var{ls}
    \\[3mm]
          &\texttt{add2HopTuples ::\texttt{\ }$\typeIP \times \typePOW{\typeL} \times \typePOW{\typeNT} \times \typeIP \times \typePOW{\typeIP \times \typeSTATUS} \times \typeTIME \rightarrow \typePOW{\typeNT}$}\\
          &\texttt{add2HopTuples(ip,ls,2hs,moip,statuses,now)} \equiv\\
          &\IF   \exists \var{lt} \in \var{ls}.\ \var{L\_oip(lt)} = \var{moip} \wedge \var{L\_status(lt,now)} = \var{SYMMETRIC} \\
          &\THEN \var{2hs} \cup \setl \var{(moip,x_1,-\infty,\infty,\infty)}\ |\ \var{x_{1..2}} \in \var{statuses} \wedge \var{x_1} \neq \var{ip} \wedge \var{x_2} = \var{SYMMETRIC}\ \wedge \\[-.4em]
&\hphantom{\THEN \var{2hs} \cup \setl \var{(moip,x_1,-\infty,\infty,\infty)}\ |\  }
           \forall \var{n2} \in \var{2hs}.\ \var{N2\_1h\_oip(n2)} \neq \var{moip} \vee \var{N2\_2h\_oip(n2)} \neq \var{x_1} \setr \\[-.3em]
          &\ELSE \var{2hs}
    \\[3mm]
          &\texttt{update2HopInMetrics ::\texttt{\ }$\typePOW{\typeL} \times \typePOW{\typeNT} \times \typeIP \times \typePOW{\typeIP \times \typeMETRIC} \times \typeTIME \rightarrow \typePOW{\typeNT}$}\\
          &\texttt{update2HopInMetrics(ls,2hs,moip,in\_metrics,now)} \equiv\\
          &\IF   \exists \var{lt} \in \var{ls}.\ \var{L\_oip(lt)} = \var{moip} \wedge \var{L\_status(lt,now)} = \var{SYMMETRIC}\\
          &\THEN \setl \var{n2} \in \var{2hs}\ |\ \var{N2\_1h\_oip(n2)} \neq \var{moip} \vee \forall \var{x_{1..2}} \in \var{in\_metrics}.\ \var{x_1} \neq \var{N2\_2h\_oip(n2)} \setr\ \cup \\[-.2em]
&\hphantom{\THEN}
                 \setl \var{n2_{1..5}}\updl \var{n2_4} := \var{x_2} \updr\ |\ \var{n2_{1..5}} \in \var{2hs} \wedge \var{N2\_1h\_oip(n2_{1..5})} = \var{moip}\ \wedge \\[-.4em]
&\hphantom{\THEN \setl \var{n2_{1..5}}\updl \var{n2_4} := \var{x_2} \updr\ |\ }
                       \var{x_{1..2}} \in \var{in\_metrics} \wedge \var{x_1} = \var{N2\_2h\_oip(n2_{1..5})}\setr \\[-.3em]
          &\ELSE \var{2hs}
    \\[3mm]
          &\texttt{update2HopOutMetrics ::\texttt{\ }$\typePOW{\typeL} \times \typePOW{\typeNT} \times \typeIP \times \typePOW{\typeIP \times \typeMETRIC} \times \typeTIME \rightarrow \typePOW{\typeNT}$}\\
          &\texttt{update2HopOutMetrics(ls,2hs,moip,out\_metrics,now)} \equiv\\
          &\IF   \exists \var{lt} \in \var{ls}.\ \var{L\_oip(lt)} = \var{moip} \wedge \var{L\_status(lt,now)} = \var{SYMMETRIC} \\
          &\THEN \setl \var{n2} \in \var{2hs}\ |\ \var{N2\_1h\_oip(n2)} \neq \var{moip} \vee \forall \var{x_{1..2}} \in \var{out\_metrics}.\ \var{x_1} \neq \var{N2\_2h\_oip(n2)} \setr\ \cup \\[-.2em]
&\hphantom{\THEN}
                 \setl \var{n2_{1..5}}\updl \var{n2_5} := \var{x_2} \updr\ |\ \var{n2_{1..5}} \in \var{2hs} \wedge \var{N2\_1h\_oip(n2_{1..5})} = \var{moip}\ \wedge \\[-.4em]
&\hphantom{\THEN \setl \var{n2_{1..5}}\updl \var{n2_5} := \var{x_2} \updr\ |\ }
                       \var{x_{1..2}} \in \var{out\_metrics} \wedge \var{x_1} = \var{N2\_2h\_oip(n2_{1..5})}\setr \\[-.3em]
          &\ELSE \var{2hs}
\end{align*}
\begin{align*}
          &\texttt{update2HopTime ::\texttt{\ }$\typeIP \times \typePOW{\typeL} \times \typePOW{\typeNT} \times \typeIP \times \typeTIME \times \typePOW{\typeIP \times \typeSTATUS} \times \typeTIME \rightarrow \typePOW{\typeNT}$}\\
          &\texttt{update2HopTime(ip,ls,2hs,moip,vtime,statuses,now)} \equiv\\
          &\IF   \exists \var{lt} \in \var{ls}.\ \var{L\_oip(lt)} = \var{moip} \wedge \var{L\_status(lt,now)} = \var{SYMMETRIC} \\
          &\THEN \setl \var{n2} \in \var{2hs}\ |\ \var{N2\_1h\_oip(n2)} \neq \var{moip} \vee \forall \var{x_{1..2}} \in \var{statuses}.\ \var{x_1} \neq \var{N2\_2h\_oip(n2)} \setr\ \cup \\
&\hphantom{\THEN}
                 \setl \var{n2_{1..5}}\updl \var{n2_3} := \var{now+vtime} \updr\ |\ \var{n2_{1..5}} \in \var{2hs} \wedge
                       \var{N2\_1h\_oip(n2_{1..5})} = \var{moip}\ \wedge \\[-.2em]
&\hphantom{\THEN \setl \var{n2_{1..5}}\updl \var{n2_3} := \var{now+vtime} \updr\ |\ }
                 \exists \var{x_{1..2}} \in \var{statuses}.\ 
                       \var{N2\_2h\_oip(n2_{1..5})} = \var{x_1}\ \wedge \\[-.3em]
&\hphantom{\THEN \setl \var{n2_{1..5}}\updl \var{n2_3} := \var{now+vtime} \updr\ |\ \exists \var{x_{1..2}} \in \var{statuses}.\ }
                       \var{x_1} \neq \var{ip}\ \wedge \var{x_2} = \var{SYMMETRIC} \setr \\[-.3em]
          &\ELSE \var{2hs}
\end{align*}

Updates to the interface information base may also be performed to preserve its consistency.
For one, expired tuples should be purged from the link set and 2-hop set,
which we achieve using the functions \texttt{purgeLinkSet} and \texttt{purge2HopSet}.
Moreover, we must recalculate and update our sets of MPRs when they no longer satisfy
the properties required of them.
First, we non-deterministically pick sets of flooding and routing mprs based on sections 18.3, 18.4 and 18.5 of RFC 7181 \cite{rfc7181}.
We then update the link set in \texttt{updateFMPRs} and \texttt{updateRMPRs} if the condition for recalculation holds.
\begin{align*}
    &\texttt{purgeLinkSet ::\texttt{\ }$\typePOW{\typeL} \times \typeTIME \rightarrow \typePOW{\typeL}$}\\
    &\texttt{purgeLinkSet(ls,now)} \equiv \\
    &\texttt{\ \ } \setl \var{lt}\ |\ \var{lt} \in \var{ls} \wedge \var{L\_time(lt)} > \var{now} \wedge \var{L\_status(lt,now)} = \var{SYMMETRIC} \setr\ \cup \\
    &\texttt{\ \ } \setl \var{lt_{1..10}} \updl \var{lt_5,lt_6,lt_7,lt_8} := \FALSE \updr \ |\ \var{lt_{1..10}} \in \var{ls} \wedge \var{L\_time(lt_{1..10})} > \var{now}\ \wedge\\[-.4em]
    &\hphantom{\texttt{\ \ } \setl \var{lt_{1..10}} \updl \var{lt_5,lt_6,lt_7,lt_8} := \FALSE \updr \ |\ }
     \var{L\_status(lt_{1..10},now)} \neq \var{SYMMETRIC} \setr
    \\[3mm]
    &\texttt{purge2HopSet ::\texttt{\ }$\typePOW{\typeL} \times \typePOW{\typeNT} \times \typeTIME \rightarrow \typePOW{\typeNT}$}\\
    &\texttt{purge2HopSet(ls,2hs,now)} \equiv \\
    &\texttt{\ \ } \setl \var{n2} \in \var{2hs}\ |\ \var{N2\_time(n2)} > \var{now} \wedge \exists \var{lt} \in \var{ls}.\ \var{L\_oip(lt)} = \var{N2\_1h\_oip(n2)}\ \wedge\\[-.4em]
    &\hphantom{\texttt{\ \ } \setl \var{n2} \in \var{2hs}\ |\ \var{N2\_time(n2)} > \var{now} \wedge \exists \var{lt} \in \var{ls}.\ }
                                                                                                                       \var{L\_status(lt,now)} = \var{SYMMETRIC} \setr
    \\[3mm]
          &\texttt{validFMPRs ::\texttt{\ }$\typePOW{\typeL} \times \typePOW{\typeNT} \times \typeTIME \rightarrow \typePOW{\typePOW{\typeL}}$}\\
          &\texttt{validFMPRs(ls,2hs,now)} \equiv\\
          &\LETand \var{N1} = \setl \var{lt} \in \var{ls}\ |\ \var{L\_status(lt,now)} = \var{SYMMETRIC} \setr \\
          &\AND \var{N2} = \setl \var{n2} \in \var{2hs}\ |\ \exists \var{lt} \in \var{N1}.\ \var{L\_oip(lt)} = \var{N2\_1h\_oip(n2)} \setr \\
          &\AND \var{d(ip,S)} = 
          \var{min}\big( \setl \infty \setr \cup
                         \setl 1\ |\ \exists \var{x} \in \var{S}.\ \var{L\_oip(x)} = \var{ip} \setr\ \cup \\
&\hphantom{\AND \var{d(ip,S)} = \var{min}\big(}
                         \setl 2\ |\ \exists \var{x} \in \var{S},\ \var{y} \in \var{N2}.\ 
                       \var{N2\_2h\_oip(y)} = \var{ip}\ \wedge \\[-.4em]
&\hphantom{\AND \var{d(ip,S)} = \var{min}\big(\setl 2\ |\ \exists \var{x} \in \var{S},\ \var{y} \in \var{N2}.\ }
                        \var{L\_oip(x)} = \var{N2\_1h\_oip(y)} \setr \big) \\
          &\INand \setl \var{M} \subseteq \var{N1}\ |\ \forall \var{y} \in \var{N2}.\ \var{d(N2\_2h\_oip(y),M)} = \var{d(N2\_2h\_oip(y),N1)} \setr
    \\[3mm]
          &\texttt{updateFMPRs ::\texttt{\ }$\typePOW{\typeL} \times \typePOW{\typeNT} \times \typeTIME \times \typePOW{\typeL} \rightarrow \typePOW{\typeL}$}\\
          &\texttt{updateFMPRs(ls,2hs,now,fmprs)} \equiv\\
          &\IF \setl \var{lt} \in \var{ls}\ |\ \var{L\_fmpr(lt)} \setr \notin \var{validFMPRs(ls,2hs,now)} \\
          &\THEN \setl \var{lt_{1..10}} \updl \var{lt_5} := \FALSE \updr\ |\ \var{lt_{1..10}} \in \var{ls} \wedge \var{lt_{1..10}} \notin \var{fmprs} \setr\ \cup \\
          &\hphantom{\THEN} \setl \var{lt_{1..10}} \updl \var{lt_5} := \TRUE \updr\ |\ \var{lt_{1..10}} \in \var{ls} \wedge \var{lt_{1..10}} \in \var{fmprs} \setr\ \cup \\
          &\ELSE \var{ls}
\end{align*}
\begin{align*}
          &\texttt{validRMPRs ::\texttt{\ }$\typePOW{\typeL} \times \typePOW{\typeNT} \times \typeTIME \rightarrow \typePOW{\typePOW{\typeL}}$}\\
          &\texttt{validRMPRs(ls,2hs,now)} \equiv\\
          &\LETand \var{N1} = \setl \var{lt} \in \var{ls}\ |\ \var{L\_status(lt,now)} = \var{SYMMETRIC} \setr \\
          &\AND \var{N2} = \setl \var{n2} \in \var{2hs}\ |\ \exists \var{lt} \in \var{N1}.\ \var{L\_oip(lt)} = \var{N2\_1h\_oip(n2)} \setr \\
          &\AND \var{d1(x)} = \var{L\_in\_metric(x)}\\
          &\AND \var{d2(y)} = \mbox{\color{red}\sout{\color{black}\var{N2\_out\_metric(y)}}} ~~\var{N2\_in\_metric(y)}\ \footnotemark 
          \\
          &\AND \var{d(ip,S)} = 
          \var{min}\big( \setl \infty \setr \cup
                         \setl \var{d1(x)}\ |\ \var{x} \in \var{S} \wedge \var{L\_oip(x)} = \var{ip} \setr\ \cup \\
&\hphantom{\AND \var{d(ip,S)} = \var{min}\big(}
                         \setl \var{d1(x)} + \var{d2(y)}\ |\ \var{x} \in \var{S} \wedge \var{y} \in \var{N2} \wedge
                       \var{N2\_2h\_oip(y)} = \var{ip} \\[-.4em]
&\hphantom{\AND \var{d(ip,S)} = \var{min}\big(\setl \var{d1(x)} + \var{d2(y)}\ |\ \var{x} \in \var{S} \wedge \var{y} \in \var{N2}}
                       \wedge \var{L\_oip(x)} = \var{N2\_1h\_oip(y)} \setr \big) \\
          &\INand \setl \var{M} \subseteq \var{N1}\ |\ \forall \var{y} \in \var{N2}.\ \var{d(N2\_2h\_oip(y),M)} = \var{d(N2\_2h\_oip(y),N1)} \setr
          \\[3mm]
          &\texttt{updateRMPRs ::\texttt{\ }$\typePOW{\typeL} \times \typePOW{\typeNT} \times \typeTIME \times \typePOW{\typeL} \rightarrow \typePOW{\typeL}$}\\
          &\texttt{updateRMPRs(ls,2hs,now,rmprs)} \equiv\\
          &\IF \setl \var{lt} \in \var{ls}\ |\ \var{L\_rmpr(lt)} \setr \notin \var{validRMPRs(ls,2hs,now)} \\
          &\THEN \setl \var{lt_{1..10}} \updl \var{lt_6} := \FALSE \updr\ |\ \var{lt_{1..10}} \in \var{ls} \wedge \var{lt_{1..10}} \notin \var{rmprs} \setr\ \cup \\
          &\hphantom{\THEN} \setl \var{lt_{1..10}} \updl \var{lt_6} := \TRUE \updr\ |\ \var{lt_{1..10}} \in \var{ls} \wedge \var{lt_{1..10}} \in \var{rmprs} \setr\ \cup \\
          &\ELSE \var{ls}
\end{align*}
\footnotetext{See \autoref{sec:bug}.}
\vspace{-1em}

\subsubsection*{Topology Information Base}

The \emph{topology information base} records information received via TC messages,
and in particular the links advertised in such messages.
It also maintains a routing set that consists of shortest routes to reachable destinations.
\begin{align*}
    &\texttt{type AR = $\typeIP \times \typeSQN \times \typeTIME$}\\
    &\texttt{type TR = $\typeIP \times \typeIP \times \typeTIME \times \typeMETRIC$}\\
    &\texttt{type R\ \  = $\typeIP \times \typeIP \times \typeMETRIC$}
\end{align*}
An entry of type \texttt{AR} in the advertising remote router set consists of
\begin{enumerate*}[label=(\emph{\roman*}), before=\unskip{: }, itemjoin={{; }}, itemjoin*={{, and }}]
    \item an originator address identifying a router from which a TC message was recently received
    \item an advertising neighbour sequence number identifying the most recent advertised information received from the originating router
    \item a validity time after which the tuple should be removed.
\end{enumerate*}
Elements of the advertising router tuples are accessed via functions of the form \var{AR\_*}.
\begin{longtable}{l@{\qquad\qquad} l@{\qquad\qquad} l}
$\begin{aligned}
          &\texttt{AR\_oip ::\texttt{\ }$\typeAR \rightarrow \typeIP$}\\
          &\texttt{AR\_oip($\var{ar_{1..3}}$)} \equiv \var{ar_1}
\end{aligned}$
&
$\begin{aligned}
          &\texttt{AR\_sqn ::\texttt{\ }$\typeAR \rightarrow \typeSQN$}\\
          &\texttt{AR\_sqn($\var{ar_{1..3}}$)} \equiv \var{ar_2}
\end{aligned}$
&
$\begin{aligned}
          &\texttt{AR\_time ::\texttt{\ }$\typeAR \rightarrow \tTIME$}\\
          &\texttt{AR\_time($\var{ar_{1..3}}$)} \equiv \var{ar_3}
\end{aligned}$
\end{longtable}

An entry of type \texttt{TR} in the router topology set consists of
\begin{enumerate*}[label=(\emph{\roman*}), before=\unskip{: }, itemjoin={{; }}, itemjoin*={{, and }}]
    \item the originator address of an advertising router
    \item the originator address of a destination router which can be reached directly via the advertising router
    \item a validity time after which the tuple should be removed
    \item a link metric from the advertising router to the destination router.
\end{enumerate*}

An entry of type \texttt{R} in the routing set consists of
\begin{enumerate*}[label=(\emph{\roman*}), before=\unskip{: }, itemjoin={{; }}, itemjoin*={{, and }}]
    \item the destination address of a reachable router
    \item the neighbour via which the router can be reached
    \item the cost of the path to the destination via the neighbour.
\end{enumerate*}

When a new TC message is processed, we store new tuples of type \typeAR\ for the advertising router and type \typeTR\ for the links advertised in the message.
\begin{align*}
          &\texttt{updateAdvertisingRouters ::\texttt{\ }$\typePOW{\typeAR} \times \typeIP \times \typeSQN \times \typeTIME \times \typeTIME \rightarrow \typePOW{\typeAR}$}\\
          &\texttt{updateAdvertisingRouters(arrs,moip,mansn,vtime,now)} \equiv \\
          &\texttt{\ \ } \setl \var{ar} \in \var{arrs}\ |\ \var{AR\_oip(ar)} \neq \var{moip} \setr \cup \setl \var{(moip,mansn,now+vtime)} \setr
          \\[3mm]
          &\texttt{updateRouterTopology ::\texttt{\ }$\typeIP \times \typePOW{\typeTR} \times \typeIP \times \typeTIME \times \typePOW{\typeIP \times \typeMETRIC} \times \typeTIME \rightarrow \typePOW{\typeTR}$}\\
          &\texttt{updateRouterTopology(ip,rts,moip,vtime,dests,now)} \equiv\\
          &\texttt{\ \ } \setl \var{tr_{1..4}} \in \var{rts}\ |\ \var{tr_1} \neq \var{moip} \setr \cup \setl \var{(moip,d_1,now+vtime,d_2)}\ |\ \var{d_{1..2}} \in \var{dests} \wedge \var{d_1} \neq \var{ip} \setr
\end{align*}

As with the interface information base, the topology information base must be kept consistent. We have two functions which remove expired tuples from the advertised remote router set and router topology set respectively. We also have a function called \texttt{incrementANSN} which records changes to the set of advertised links by incrementing the router's advertising neighbour sequence number.
\begin{align*}
    &\texttt{purgeAdvertisingRouters ::\texttt{\ }$\typePOW{\typeAR} \times \typeTIME \rightarrow \typePOW{\typeAR}$}\\
    &\texttt{purgeAdvertisingRouters(arrs,now)} \equiv \setl \var{ar} \in \var{arrs}\ |\ \var{AR\_time(ar)} > \var{now} \setr
          \\[3mm]
    &\texttt{purgeRouterTopology ::\texttt{\ }$\typePOW{\typeTR} \times \typeTIME \rightarrow \typePOW{\typeTR}$}\\
    &\texttt{purgeRouterTopology(rts,now)} \equiv \setl \var{tr_{1..4}} \in \var{rts}\ |\ \var{tr_3} > \var{now} \setr
          \\[3mm]
          &\texttt{incrementANSN ::\texttt{\ }$\typePOW{\typeL} \times \typePOW{\typeL} \times \typeSQN \rightarrow \typeSQN$}\\
          &\texttt{incrementANSN(ls,prev\_ls,ansn)} \equiv \\
          &\IF \setl \var{L\_oip(lt)}\ |\ \var{lt} \in \var{ls} \wedge \var{L\_rmpr\_selector(lt)} \setr \neq \\
&\hphantom{\IF} \setl \var{L\_oip(lt)}\ |\ \var{lt} \in \var{prev\_ls} \wedge \var{L\_rmpr\_selector(lt)} \setr \\
          &\THEN \var{ansn + 1}
           \texttt{\ \bf else\ } \var{ansn}
\end{align*}

Finally, we have a function which calculates a router's optimal routing sets
and another which updates the routing set to an optimal set if the current set is not optimal.
\begin{align*}
          &\texttt{optimalRoutingSets ::\texttt{\ }$\typeIP \times \typePOW{\typeL} \times \typePOW{\typeTR} \times \typeTIME \rightarrow \typePOW{\typePOW{\typeR}}$}\\[-.2em]
          &\texttt{optimalRoutingSets(ip,ls,rts,now)} \equiv\\
          &\LETand \var{links} = \setl \var{(tr_1,tr_2,tr_4)}\ |\ \var{tr_{1..4}} \in \var{rts} \setr\ \cup\\
          &\hphantom{\LETand \var{paths} =\ }
           \setl \var{(ip,L\_oip(lt),L\_out\_metric(lt))}\ |\ \var{lt} \in \var{ls} \wedge \var{status(lt,now)} = \var{SYMMETRIC} \setr\\
          &\AND \var{routes} = \\[-1em]
          &\hphantom{\AND}\setl \var{(d_n,d_1,m)}\ |\  \var{(s_1,d_1,m_1)}...\var{(s_n,d_n,m_n)} \in \var{links}^+ \wedge 
           \var{s_1} = \var{ip} \wedge \bigwedge_{i=1}^{n-1} \var{d_i} = \var{s_{i+1}} \wedge \var{m} = \sum_{i=1}^{n}\var{m_i} \setr\\
          &\AND \var{shortest\_routes} = \setl \var{(d,s,m)} \in \var{routes}\ |\ \forall \var{(d',s',m')} \in \var{routes}.\ \ \var{d} = \var{d'} \Longrightarrow \var{m} \le \var{m'} \setr \\
          &\INand \setl \var{rs'} \subseteq \var{shortest\_routes}\ |\ (\forall \var{(d,s,m)} \in \var{routes}.\ \exists \var{(d',s',m')} \in \var{rs'}.\ \var{d} = \var{d'})\ \wedge \\[-.4em]
          &\hphantom{\INand \setl \var{rs'} \subseteq \var{shortest\_routes}\ |\ }
           (\forall \var{(d,s,m)},\var{(d',s',m')} \in \var{rs'}.\ \var{d} = \var{d'} \Longrightarrow \var{s} = \var{s'} \wedge \var{m} = \var{m'}) \setr
\end{align*}

\pagebreak

\noindent\begin{align*}
          &\texttt{updateRoutingSet ::\texttt{\ }$\typeIP \times \typePOW{\typeL} \times \typePOW{\typeTR} \times \typeTIME \times \typePOW{\typeR} \times \typePOW{\typeR} \rightarrow \typePOW{\typeR}$}\\
          &\texttt{updateRoutingSet(ip,ls,rts,now,rs,rs$'$)} \equiv\\
          &\texttt{\ \ \bf if\ \ } \var{rs} \notin \var{optimalRoutingSets(ip,ls,rts,now)} 
           \texttt{\ \bf then\ } \var{rs'}
           \texttt{\ \bf else\ } \var{rs}
\end{align*}

\subsubsection*{Received Message Information Base}

The \emph{received message information base} records all TC messages that have been processed and received by the router. Messages are uniquely identified by their originator address and sequence number, so we store both pieces of information for when we make future processing and forwarding decisions.
\begin{align*}
    &\texttt{type P\ \ = $\typeIP \times \typeSQN$}\\
    &\texttt{type RX = $\typeIP \times \typeSQN$}
\end{align*}
As with the other information bases, we define some basic functions for extracting information from tuples.
\begin{longtable}{l l l l}
$\begin{aligned}
          &\texttt{P\_oip ::\texttt{\ }$\typeP \rightarrow \typeIP$}\\
          &\texttt{P\_oip($\var{p_{1..2}}$)} \equiv \var{p_1}
\end{aligned}$
&
$\begin{aligned}
          &\texttt{P\_sqn ::\texttt{\ }$\typeP \rightarrow \typeSQN$}\\
          &\texttt{P\_sqn($\var{p_{1..2}}$)} \equiv \var{p_2}
\end{aligned}$
&
$\begin{aligned}
          &\texttt{RX\_oip ::\texttt{\ }$\typeRX \rightarrow \typeIP$}\\
          &\texttt{RX\_oip($\var{rx_{1..2}}$)} \equiv \var{rx_1}
\end{aligned}$
&
$\begin{aligned}
          &\texttt{RX\_sqn ::\texttt{\ }$\typeRX \rightarrow \typeSQN$}\\
          &\texttt{RX\_sqn($\var{rx_{1..2}}$)} \equiv \var{rx_2}
\end{aligned}$
\end{longtable}
\noindent We also provide two functions for storing new tuples in either a processed set or a received set.
\begin{align*}
          &\texttt{addProcessedTuple ::\texttt{\ }$\typePOW{\typeP} \times \typeIP \times \typeSQN \rightarrow \typePOW{\typeP}$}\\
          &\texttt{addProcessedTuple(ps,moip,msqn)} \equiv \var{ps} \cup \setl \var{(moip,msqn)} \setr
          \\[3mm]
          &\texttt{addReceivedTuple ::\texttt{\ }$\typePOW{\typeRX} \times \typeIP \times \typeSQN \rightarrow \typePOW{\typeRX}$}\\
          &\texttt{addReceivedTuple(rxs,moip,msqn)} \equiv \var{rxs} \cup \setl \var{(moip,msqn)} \setr
\end{align*}

\subsubsection*{Information Base Changes}

In our model, we localise all relevant information base updates to a single process.
The condition to call the process and trigger these updates, given by the \texttt{updatesPending} function, asserts that the updates would modify the protocol's information bases and implies that they are currently inconsistent.
\begin{align*}
          &\texttt{updatesPending ::\texttt{\ }$\typePOW{\typeL} \times \typePOW{\typeNT} \times
    \typePOW{\typeAR} \times \typePOW{\typeTR} \times \typePOW{\typeR} \times \typeSQN \times
    \typePOW{\typeL} \times \typeIP \times \typeTIME \rightarrow \typeBOOL$}\\
          &\texttt{updatesPending(ls,2hs,arrs,rts,rs,ansn,prev\_ls,ip,now)} \equiv \\
          &\texttt{\ \ }\var{ls} \neq \var{purgeLinkSet(ls,now)}\ \vee \\
          &\texttt{\ \ }\var{2hs} \neq \var{purge2HopSet(ls,2hs,now)}\ \vee \\
          &\texttt{\ \ }\var{arrs} \neq \var{purgeAdvertisingRouters(arrs,now)}\ \vee \\
          &\texttt{\ \ }\var{rts} \neq \var{purgeRouterTopology(rts,now)}\ \vee \\
          &\texttt{\ \ }\setl \var{lt} \in \var{ls}\ |\ \var{L\_fmpr(lt)} \setr \notin \var{validFMPRs(ls,2hs,now)}\ \vee \\
          &\texttt{\ \ }\setl \var{lt} \in \var{ls}\ |\ \var{L\_rmpr(lt)} \setr \notin \var{validRMPRs(ls,2hs,now)}\ \vee \\
          &\texttt{\ \ }\var{ansn} \neq \var{incrementANSN(ls,prev\_ls,ansn)}\ \vee \\
          &\texttt{\ \ }\var{rs} \notin \var{optimalRoutingSets(ip,ls,rts,now)}
\end{align*}

\subsubsection*{Process State}

\setcounter{table}{1}

Our protocol maintains its state in a list of variables. The main variables are summarised in \autoref{table:variables}.
We make a distinction between those variables that are modified during the protocol's execution and those that are not.
For the former, we use the shorthand $\sigma$
to represent a comma-separated list of these variables.
Similarly, we use $\Gamma$ for the latter.
We also maintain a variable \texttt{queue} of type \typeList{\typeMSG}\ in our input queue process and a variable \texttt{msg} of type \typeMsg\ when processing or forwarding a message.
\begin{table}[!ht]
\caption{The variables constituting the protocol's state\label{table:variables}}
\vspace{3pt}
\tabulinesep=4pt
\centering
\begin{tabu} to .98\textwidth{@{}r@{}|X[6,c]|X[3.8,c]|X[19.5,c]|}
    \cline{2-4}
   & \textbf{Name} & \textbf{Type} & \textbf{Description} \\
    \cline{2-4}
   \multirow{15}{*}[55pt]{$\sigma\left\{\rule{0pt}{169pt}\right.$}
   & \var{ls} & \typePOW{\typeL} & Link set maintaining information about 1-hop neighbours and their statuses \\
   \cline{2-4}
   & \var{2hs} & \typePOW{\typeNT} & 2-hop set maintaining information about 2-hop neighbours \\
    \cline{2-4}
   & \var{arrs} & \typePOW{\typeAR} & Advertising remote router set containing information about routers which have advertised links \\
    \cline{2-4}
   & \var{rts} & \typePOW{\typeTR} & Router topology set containing advertised links \\
    \cline{2-4}
   & \var{rs} & \typePOW{\typeR} & Routing set containing shortest known routes \\
    \cline{2-4}
   & \var{ps} & \typePOW{\typeP} & Processed set identifying processed TC messages \\
    \cline{2-4}
   & \var{rxs} & \typePOW{\typeRX} & Received set identifying TC messages received and considered for forwarding \\
    \cline{2-4}
   & \var{pkt} & \typeList{\typeMsg} & List of messages requiring sending, such as those generated by the router or forwarded by it \\
    \cline{2-4}
   & \var{hello\_time} & \typeTIME & Time when next HELLO message must be added to \var{pkt}\\
    \cline{2-4}
   & \var{tc\_time} & \typeTIME & Time when next TC message must be added to \var{pkt}\\
    \cline{2-4}
   & \var{send\_time} & \typeTIME & Time when \var{pkt} must be broadcast\\
    \cline{2-4}
   & \var{mqueue} & \typeList{\typeMsg} & Queue of to-be-processed messages \\
    \cline{2-4}
   & \var{sqn} & \typeSQN & Sequence number identifying a TC message \\
    \cline{2-4}
   & \var{ansn} & \typeSQN & Advertising neighbour sequence number included in TC messages to indicate how recent the advertised contents are \\
    \cline{2-4}
   & \var{prev\_ls} & \typePOW{\typeL} & Previous link set used to check for updates \\
    \hline
    \multirow{8}{*}[1pt]{$\Gamma\left\{\rule{0pt}{74pt}\right.$}
   & \var{ip} & \typeIP & Address of the router \\
    \cline{2-4}
   & \var{hp\_maxjitter} & \typeTIME & Maximum jitter time for HELLO messages \\
    \cline{2-4}
   & \var{tp\_maxjitter} & \typeTIME & Maximum jitter time for TC messages \\
    \cline{2-4}
   & \var{h\_hold\_time} & \typeTIME & Validity time for generated HELLO messages \\
    \cline{2-4}
   & \var{t\_hold\_time} & \typeTIME & Validity time for generated TC messages \\
    \cline{2-4}
   & \var{l\_hold\_time} & \typeTIME & Length that lost links should be kept for \\
    \cline{2-4}
   & \var{hello\_interval} & \typeTIME & Period between HELLO message transmissions \\
    \cline{2-4}
   & \var{tc\_interval} & \typeTIME & Period between TC message transmissions \\
    \cline{2-4}
\end{tabu}
\end{table}

\subsection{\tawn-Specification of OLSRv2\label{app:model}}
In this section, we present our T-AWN model of OLSRv2.
The model consists of five main processes implementing the OLSRv2 specification
and a queue process to receive packets from other routers:

\begin{description}[nolistsep]
    \item [\texttt{OLSR} (\autoref{proc:olsrapp}):]
The \texttt{OLSR} process constitutes the main protocol loop. It is responsible for receiving packets from the input queue and processing these packets according to their type. It it also responsible for periodically generating new HELLO and TC messages.
    \item  [\texttt{UPDATE\_INFO} (\autoref{proc:update}):]
The \texttt{UPDATE\_INFO} process ensures that the protocol's information bases remain consistent with certain constraints.
    \item  [\texttt{PROCESS\_HELLO} (\autoref{proc:hello}):]
The \texttt{PROCESS\_HELLO} process is responsible for recording information obtained through HELLO messages in the relevant information bases.
    \item  [\texttt{PROCESS\_TC}, (\autoref{proc:tc}):]
The \texttt{PROCESS\_TC} process is responsible for recording information obtained through TC messages in the relevant information bases.
    \item  [\texttt{FORWARD\_TC}, (\autoref{proc:forward}):]
The \texttt{FORWARD\_TC} process forwards TC messages received from the router's flooding MPR selectors, subject to a few side conditions.
    \item  [\texttt{QUEUE}, (\autoref{proc:queue}):]
The \texttt{QUEUE} process receives packets from other routers in the network and delivers them to the \texttt{OLSR} process.
\end{description}

For clarity, we use three pieces of syntactic sugar in our guards.
The first, \texttt{otherwise}, stands for the negation of the previous guard.
The second, \texttt{Let}, denotes a guard used as a non-deterministic assignment but has no meaning otherwise.
The third, \var{Updated}, is shorthand for the expression\\
       \centerline{$\neg\var{updatesPending(ls,2hs,arrs,rts,rs,ansn,prev\_ls,ip,now)}\ .$}

\subsubsection{The Main Routine}
\label{app:main}

The main OLSR routine performs a number of different roles.
The most basic of these is receiving a packet from the input queue, which occurs in the block on lines \hyperref[proc:olsrapp]{1-3}.
Packets of type \texttt{MSG} are simply lists of HELLO and TC messages, so we concatenate received packets with an existing queue of to-be-processed messages.
When the protocol is ready to process a HELLO or TC message,
the choice on line \hyperref[proc:olsrapp]{8} is taken,
with \var{msg} and \var{msgs} assigned to the head and tail of \var{mqueue}
respectively by the guard. Note that this guard contains an \texttt{Updated} conjunct, which asserts that the information bases have already been updated by the block on lines \hyperref[proc:olsrapp]{5-6}. Once \var{msg} is assigned to the element at the head of the queue,
the ensuing assignment statement assigns \texttt{mqueue} to its tail \var{msgs}. The guards on lines \hyperref[proc:olsrapp]{9} and \hyperref[proc:olsrapp]{12} then ensure that \texttt{msg} is processed according to its type,
be that a HELLO message or a TC message.

\setcounter{algocf}{0}
\stepcounter{counterfix}

\begin{algorithm}
{\small
\SetAlgoNoLine
\caption{The main OLSR process\label{proc:olsrapp}}
\vspace{1pt}
\OLSRDEF{\ensuremath{\sigma,\Gamma}}
\vspace{1pt}
\tcc{Receive a packet (i.e.\ a list of messages) from the queue process}
\procRECV{\var{msgs}}{
    \procASSN{\var{mqueue} := \var{concat(mqueue,msgs)}}{}
    \procDEFN{OLSR}{\sigma,\Gamma}
}
\vspace{-.15em}
\procCHOICE
\tcc{Execute pending updates to relevant information bases}
\procGUARD{\neg\var{Updated}}{
    \procDEFN{UPDATE\_INFO}{\sigma,\Gamma}\\
}
\vspace{-.15em}
\procCHOICE
\tcc{Process a received message}
\procGGUARD{\var{Updated}\ \wedge\ \var{send\_time} \neq \var{now} \wedge\ \var{mqueue} = (\var{msg}:\var{msgs})}{}
\procASSN{\var{mqueue} := \var{msgs}}{
    \tcc{Process a received HELLO message}
    \procGUARD{\var{isHELLO(msg)}}{
    \procDEFN{PROCESS\_HELLO}{\sigma,\Gamma,\var{msg}}\\
    }
    \vspace{-.15em}
    \procCHOICE
    \tcc{Process a received TC message}
    \procGUARD{\var{isTC(msg)}}{
    \procDEFN{PROCESS\_TC}{\sigma,\Gamma,\var{msg}}\\
    }
}
\vspace{-.15em}
\procCHOICE
\tcc{Time to generate a HELLO message}
\procGUARD{\var{Updated}\ \wedge\ \var{send\_time} \neq \var{now} \wedge\ \var{hello\_time} - \var{hp\_maxjitter} \le \var{now} \le \var{hello\_time}}{
    \tcc{Add the message to the current packet}
    \procASSN{\var{pkt} := \var{append(pkt,newHELLO(ip,h\_hold\_time,ls,now))}}{}
    \tcc{Set relevant timers}
    \procASSN{\var{hello\_time} := \var{now} + \var{hello\_interval}}{}
    \procASSN{\var{send\_time} := \var{now}+1}{}
    \procDEFN{OLSR}{\sigma,\Gamma}
}
\vspace{-.15em}
\procCHOICE
\tcc{Time to generate a TC message}
\procGUARD{\var{Updated}\ \wedge\ \var{send\_time} \neq \var{now} \wedge\ \var{tc\_time} - \var{tp\_maxjitter} \le \texttt{now} \le \var{tc\_time}}{
    \tcc{Add the message to the current packet}
    \procASSN{\var{pkt} := \var{append(pkt,newTC(ip,t\_hold\_time,sqn,ansn,ls,now))}}{}
    \tcc{Increment the sequence number}
    \procASSN{\var{sqn} := \var{sqn} + 1}{}
    \tcc{Set relevant timers}
    \procASSN{\var{tc\_time} := \var{now} + \var{tc\_interval}}{}
    \procASSN{\var{send\_time} := \var{now}+1}{}
    \procDEFN{OLSR}{\sigma,\Gamma}
}
\vspace{-.15em}
\procCHOICE
\tcc{Broadcast the accumulated packet}
\procGUARD{\var{Updated}\ \wedge\ \texttt{send\_time} = \texttt{now}}{
    \procASSN{\texttt{send\_time} := \infty}{}
    \procBC{\var{pkt}}{}
    \procASSN{\var{pkt} := [\hspace{.3em}]}{}
    \procDEFN{OLSR}{\sigma,\Gamma}
}
\vspace{1pt}
}
\end{algorithm}

Aside from processing messages, a router must generate its own HELLO and TC messages periodically. The guard on line \hyperref[proc:olsrapp]{15} asserts that a new HELLO message is ready to be added to \var{pkt}, a list which accumulates all messages generated during a time tick in order to circumvent broadcasting delays. The guard is true whenever the local clock \texttt{now} enters the jitter period before the message's preparation deadline \texttt{hello\_time}. The condition $\var{now} \le \var{hello\_time}$ is redundant, as one can   prove that it is always met even when left out; it reminds us that we have not yet exceeded the deadline $\var{hello\_time}$ when line   \hyperref[proc:olsrapp]{16} is executed. A new HELLO message is generated by line \hyperref[proc:olsrapp]{16} and appended to \texttt{pkt}. Then, \texttt{hello\_time} is increased to \texttt{hello\_interval} time units away from \var{now}, and \texttt{send\_time} is set to $\var{now}+1$ so that the packet will be broadcast during the next tick. Sending a TC message in the block starting at line \hyperref[proc:olsrapp]{21} follows a similar process,
except that a sequence number \var{sqn} is included in the message and subsequently incremented.
Once all messages have been accumulated and the guard on line \hyperref[proc:olsrapp]{28} becomes true, the accumulated packet is broadcast, and both \texttt{pkt} and \texttt{send\_time} are reset to indicate no pending messages.

\subsubsection{Updating the Information Bases}\label{app:update}

Changes to the router's information bases may trigger updates such as those defined in section 13 of RFC 6130 \cite{rfc6130}. Whenever an update is triggered, the \texttt{UPDATE\_INFO} process is called to restore the consistency of these information bases. On lines \hyperref[proc:update]{1-4}, expired tuples are removed from the link set, 2-hop set, advertising remove router set and router topology set. The functions \texttt{purgeLinkSet} and \texttt{purge2HopSet} perform additional steps to maintain consistency, with \texttt{purgeLinkSet} resetting the MPR statuses of non-symmetric tuples and \texttt{purge2HopSet} removing all tuples without a symmetric neighbour. Next, a set of valid flooding MPRs and a set of valid routing MPRs are non-deterministically chosen by lines \hyperref[proc:update]{5} and \hyperref[proc:update]{7}.
The link set is updated to use these MPR sets
on lines \hyperref[proc:update]{6} and \hyperref[proc:update]{8}
iff the currently chosen MPR sets are not valid. On line \hyperref[proc:update]{9}, we increment the advertising neighbour sequence number to indicate a change in advertised information iff the set of routing MPR selectors has changed. We then assign \texttt{prev\_ls} to the current link set so that we can test for changes in the future.
Finally, the routing set is updated on lines \hyperref[proc:update]{11} and \hyperref[proc:update]{12} iff
it does not use optimal routes to all known destinations.

\begin{algorithm}
\SetAlgoNoLine
{\small
\caption{Update the information bases to maintain consistency\label{proc:update}}
\vspace{2pt}
\updateDEF{\ensuremath{\sigma,\Gamma}}
\vspace{2pt}
\tcc{Remove expired tuples, perform additional consistency checks}
\procASSN{\var{ls} := \var{purgeLinkSet(ls,now)}}{}
\procASSN{\var{2hs} := \var{purge2HopSet(ls,2hs,now)}}{}
\procASSN{\var{arrs} := \var{purgeAdvertisingRouters(arrs,now)}}{}
\procASSN{\var{rts} := \var{purgeRouterTopology(rts,now)}}{}
\tcc{Update the router's flooding and routing MPRs if necessary}
\procGUARD{\var{Let}\ \var{fmprs} \in \var{validFMPRs(ls,2hs,now)}}{}
\procASSN{\var{ls} := \var{updateFMPRs(ls,2hs,now,fmprs)}}{}
\procGUARD{\var{Let}\ \var{rmprs} \in \var{validRMPRs(ls,2hs,now)}}{}
\procASSN{\var{ls} := \var{updateRMPRs(ls,2hs,now,rmprs)}}{}
\tcc{Increment the advertising neighbour sequence number if the routing MPR selectors have changed}
\procASSN{\var{ansn} := \var{incrementANSN(ls,prev\_ls,ansn)}}{}
\procASSN{\var{prev\_ls} := \var{ls}}{}
\tcc{If the current routing set is invalid, update it}
\procGUARD{\var{Let}\ \var{rs'} \in \var{optimalRoutingSets(ip,ls,rts,now)}}{}
\procASSN{\var{rs} := \var{updateRoutingSet(ip,ls,rts,now,rs,rs')}}{}
\procDEFN{OLSR}{\sigma,\Gamma}
\vspace{2pt}
}
\end{algorithm}
\vspace{-1em}

\subsubsection{Processing a HELLO Message}\label{app:hello}

When a new HELLO message is received by a router, its link set and 2-hop set must be updated. We
perform the updates to the link set in the first half of \texttt{PROCESS\_HELLO}. First, we
non-deterministically choose an incoming link metric from the router sending the
message to the receiving router. In fact, this link metric is usually determined by the quality of the
link, as measured by the receiving router. However, the RFC does not specify how this is done, and
therefore we simply model this as a nondeterministic choice.
A new tuple using this incoming metric is then created with an originator address equal to the
sending router iff such a tuple did not already exist in the link set. The link tuple for the sending
router, existing or new, is then updated based on the contents of the message.
First, line \hyperref[proc:hello]{3} updates the outgoing metric of the tuple to the incoming
metric measured at the sending router iff this element was advertised in the message.
Next, the symmetric, heard and expiry times of the tuple are modified on lines \hyperref[proc:hello]{4}, \hyperref[proc:hello]{5} and \hyperref[proc:hello]{6}. Both the heard and expiry times will be made valid for at least the validity time of the message, whereas the symmetric time will be updated based on whether the router found itself advertised as a neighbour in the HELLO message.
The router also determines whether it has been selected as a flooding MPR or a routing MPR from the contents of the message.

In the second half of the process, the 2-hop set is updated to ensure that 2-hop tuples exist for all neighbours of the sending router.
On line \hyperref[proc:hello]{9}, new 2-hop tuples are created for neighbours besides the receiving router if they do not already exist.
Next, the metrics and validity time of these tuples are updated.
All of these functions require there to exist a symmetric link tuple for the sending router.
If not, then the assignments have no effect.

\begin{algorithm}
\SetAlgoNoLine
{\small
\caption{Process a HELLO message\label{proc:hello}}
\vspace{2pt}
\HELLODEF{\ensuremath{\sigma,\Gamma,\var{msg}}}
\vspace{2pt}
\tcc{Update the link set}
\procGUARD{\var{Let}\ \var{in\_metric} \neq \infty}{}
\procASSN{\var{ls} := \var{addLinkTuple(ls,oip(msg),vtime(msg),in\_metric,now)}}{}
\procASSN{\var{ls} := \var{updateLinkOutMetrics(ip,ls,oip(msg),inMetrics(msg))}}{}
\procASSN{\var{ls} := \var{updateSymmetricTime(ip,ls,oip(msg),vtime(msg),statuses(msg),l\_hold\_time,now)}}{}
\procASSN{\var{ls} := \var{updateHeardTime(ls,oip(msg),vtime(msg),now)}}{}
\procASSN{\var{ls} := \var{updateValidityTime(ls,oip(msg),l\_hold\_time,now)}}{}
\procASSN{\var{ls} := \var{updateFMPRSelectors(ip,ls,oip(msg),statuses(msg),mprs(msg),now)}}{}
\procASSN{\var{ls} := \var{updateRMPRSelectors(ip,ls,oip(msg),statuses(msg),mprs(msg),now)}}{}
\tcc{Update the 2-hop set}
\procASSN{\var{2hs} := \var{add2HopTuples(ip,ls,2hs,oip(msg),statuses(msg),now)}}{}
\procASSN{\var{2hs} := \var{update2HopInMetrics(ls,2hs,oip(msg),inMetrics(msg),now)}}{}
\procASSN{\var{2hs} := \var{update2HopOutMetrics(ls,2hs,oip(msg),outMetrics(msg),now)}}{}
\procASSN{\var{2hs} := \var{update2HopTime(ip,ls,2hs,oip(msg),vtime(msg),statuses(msg),now)}}{}
\procDEFN{OLSR}{\sigma,\Gamma}
\vspace{2pt}
}
\end{algorithm}
\vspace{-1em}

\subsubsection{Processing a TC Message}\label{app:tc}

TC messages must also be processed once received, subject to a few additional checks. Firstly, the message must be discarded if it was originated by the receiving router. This check is required because TC messages, unlike HELLO message, are flooded through the network and may eventually reach the router that generated them. Processing is also optional if the message was not received from a known symmetric neighbour of the router, i.e. there is no symmetric link tuple for the sending router.

If the guard on line \hyperref[proc:tc]{7} is true, then two additional checks must be performed prior to processing. Firstly, we check whether or not the message has been processed before. This is determined by the guard on line \hyperref[proc:tc]{8}, which asserts that there is a processed tuple with the same originator address and message sequence number as the received message. If there is no such tuple, we create a new one for the message and add it to the processed set. The last guard, located on line \hyperref[proc:tc]{13}, checks the advertising neighbour sequence number of the message and determines whether a message from the same router with a greater advertising neighbour sequence number has already been received. If not, we create an advertising remote router tuple and record the advertised links in the router topology set.
\vspace{2ex}

\begin{algorithm}[H]
\SetAlgoNoLine
{\small
\caption{Process a TC message\label{proc:tc}}
\vspace{2pt}
\TCDEF{\ensuremath{\sigma,\Gamma,\var{msg}}}
\vspace{2pt}
{
\tcc{If the message was originated by this router, then discard it}
\procGUARD{\var{oip(msg)} = \var{ip}}{
    \procDEFN{OLSR}{\sigma,\Gamma}
}
\procCHOICE
\tcc{If the message was not received from a known symmetric neighbour, then processing is optional}
\procGUARD{\var{oip(msg)} \neq \var{ip} \wedge \forall \var{lt} \in \var{ls}.\ \ \var{L\_oip(lt)} \neq \var{sip(msg)} \vee \var{L\_status(lt,now)} \neq \var{SYMMETRIC}}{
    \procDEFN{FORWARD\_TC}{\sigma,\Gamma,\var{msg}}
}
}
\procCHOICE
\tcc{The message was not originated by this router}
\procGUARD{\var{oip(msg)} \neq \var{ip}}{
    \tcc{If a message with the same originating router and sequence number was received previously, then do not process the message}
    \procGUARD{\exists \var{p} \in \var{ps}.\ \ \var{P\_oip(p)} = \var{oip(msg)} \wedge\ \var{P\_sqn(p)} = \var{sqn(msg)}}{
        \procDEFN{FORWARD\_TC}{\sigma,\Gamma,\var{msg}}
    }
    \procCHOICE
    \procGUARD{\var{otherwise}}{
        \tcc{Mark the message as processed}
        \procASSN{\var{ps} := \var{addProcessedTuple(ps,oip(msg),sqn(msg))}}{
            \tcc{If the advertising neighbour sequence number included in the message is out of date, then discard the message}
            \procGUARD{\exists \var{ar} \in \var{arrs}.\ \ \var{AR\_oip(ar)} = \var{oip(msg)} \wedge\ \var{AR\_sqn(ar)} > \var{ansn(msg)}}{
                \procDEFN{FORWARD\_TC}{\sigma,\Gamma,\var{msg}}
            }
            \procCHOICE
            \tcc{Process the advertised information}
            \procGUARD{\var{otherwise}}{
                \procASSN{\var{arrs} := \var{updateAdvertisingRouters(arrs,oip(msg),ansn(msg),vtime(msg),now)}}{}
                \procASSN{\var{rts} := \var{updateRouterTopology(ip,rts,oip(msg),vtime(msg),dests(msg),now)}}{}
                \procDEFN{FORWARD\_TC}{\sigma,\Gamma,\var{msg}}
            }
        }
    }
}
\vspace{2pt}
}
\end{algorithm}

\subsubsection{Forwarding a TC Message}\label{app:forward}
If a router receives a TC message,
then that message will be considered for forwarding
iff the message was not originated by the receiving router.
The guard on line \hyperref[proc:forward]{1} discards messages that were not received from a known symmetric neighbour.
The message is also discarded if the guard on line \hyperref[proc:forward]{5} is true,
indicating that the message was received previously.
Once these guards are passed, a new received tuple for the message is created to prevent it from being forwarded again in the future. The final guard on line \hyperref[proc:forward]{10} asserts that the router which last forwarded the message is a flooding MPR of this router. In this case, the message should be forwarded. The router then modifies the message's sender IP address to its own IP address and appends it to \var{pkt} on line \hyperref[proc:forward]{11}. Finally, before returning to the main protocol loop, \var{send\_time} is assigned to $\var{now} + 1$ to trigger an eventual broadcast of the packet.
\vspace{1ex}

\begin{algorithm}[H]
\SetAlgoNoLine
{\small
\caption{Forward a TC message\label{proc:forward}}
\vspace{2pt}
\FORWARDDEF{\ensuremath{\sigma,\Gamma,\var{msg}}}
\vspace{2pt}
{
\tcc{If the message was not received from a known symmetric neighbour, then discard it}
\procGUARD{\forall \var{lt} \in \var{ls}.\ \ \var{L\_oip(lt)} \neq \var{sip(msg)} \vee \var{L\_status(lt,now)} \neq \var{SYMMETRIC}}{
    \procDEFN{OLSR}{\sigma,\Gamma}
}
\procCHOICE
}
\procGUARD{\var{otherwise}}{
    \tcc{If a message with the same originator address and sequence number was received previously, do not consider the message for forwarding}
    \procGUARD{\exists \var{rx} \in \var{rxs}.\ \ \var{RX\_oip(rx)} = \var{oip(msg)} \wedge\ \var{RX\_sqn(rx)} = \var{sqn(msg)}}{
        \procDEFN{OLSR}{\sigma,\Gamma}
    }
    \procCHOICE
    \procGUARD{\var{otherwise}}{
        \tcc{Create a received tuple for the message}
        \procASSN{\var{rxs} := \var{addReceivedTuple(rxs,oip(msg),sqn(msg))}}{
            \tcc{If the message was received from a flooding MPR selector, forward the message}
            \procGUARD{\exists \var{lt} \in \var{ls}.\ \ \var{L\_oip(lt)} = \var{sip(msg)} \wedge\ \var{L\_fmpr\_selector(lt)}}{
                \procASSN{\var{pkt} := \var{append(pkt,forward(ip,msg))}}{}
                \procASSN{\var{send\_time} := \var{now+1}}{}
                \procDEFN{OLSR}{\sigma,\Gamma}
            }
            \procCHOICE
            \tcc{If the message was not received from a flooding MPR selector, do not forward it}
            \procGUARD{\var{otherwise}}{
                \procDEFN{OLSR}{\sigma,\Gamma}
            }
        }
    }
}
\vspace{2pt}
}
\end{algorithm}

\subsubsection{The Message Queue}
\label{app:queue}

The \texttt{QUEUE} process can either receive a message and append it to the \texttt{queue} variable, as on lines \hyperref[proc:queue]{1} and~\hyperref[proc:queue]{6}, or send a message to the \texttt{OLSR} process, as on line \hyperref[proc:queue]{4}. The guard on line \hyperref[proc:queue]{3} asserts that the queue is not empty and assigns the free variables \texttt{q} and \texttt{qs} to its head and tail respectively. The additional receive on line \hyperref[proc:queue]{6} is needed to prevent blocking, such as when the \texttt{OLSR} process cannot currently receive a packet but the \texttt{QUEUE} process can.

\begin{algorithm}
\SetAlgoNoLine
{\small
\caption{Message Queue\label{proc:queue}}
\vspace{2pt}
\QUEUEDEF{$\texttt{queue}$}
\vspace{2pt}
\tcc{Receive a packet from another router and append it to the queue}
\procRRECV{\var{pkt}}{}
\procDEFN{QUEUE}{\texttt{append(queue,pkt)}}\\
\procCHOICE
\tcc{Packet queue is not empty}
\procGUARD{\var{queue} = \var{(q:qs)}}{
    \tcc{Dequeue a packet and send it to the main OLSR process}
    \procSSEND{\var{q}}{}
    \procDEFN{QUEUE}{\var{qs}}\\
    \procCHOICE
    \tcc{Receive a packet from another router and append it to the queue}
    \procRRECV{\var{pkt}}{}
    \procDEFN{QUEUE}{\texttt{append(queue,pkt)}}
}
\vspace{2pt}
}
\end{algorithm}

\subsubsection{Initial State and Constraints}
\label{ssec:state}

The initial state of the system is expressed as an encapsulated parallel composition of a finite number of nodes, each of the form\\
\centerline{$\dval{ip} : (\xi,\texttt{OLSR($\sigma,\Gamma$)}\ \langle\kern -1pt \langle\ \zeta,\texttt{QUEUE(queue)}) : R\ .$}

\smallskip
\noindent In practice, we restrict the addresses $\dval{ip}$ of nodes to a finite set $\pred{IP}$
and ensure that each address in $\pred{IP}$ corresponds to exactly one node.
Then, $R \subseteq \pred{IP}$ represents the nodes currently within range of a router.
In a network of $|\pred{IP}|$ nodes with minimum broadcast duration $\var{LB}$
and maximum broadcast duration $\var{LB} + \Delta\var{B}$, we use the conjunction of the following three formulas to constrain
the initial values of $\xi$ and $\zeta$.
\begin{gather*}
\xi(\var{ls}) = \emptyset\ \wedge\ 
\xi(\var{2hs}) = \emptyset\ \wedge\
\xi(\var{arrs}) = \emptyset\ \wedge\
\xi(\var{rts}) = \emptyset\ \ \wedge\\[-.22em]
\xi(\var{rs}) = \emptyset\ \wedge\
\xi(\var{ps}) = \emptyset\ \wedge\
\xi(\var{rxs}) = \emptyset\ \wedge
\xi(\var{prev\_ls}) = \emptyset\ \wedge\\[-.22em]
\xi(\var{sqn}) = 0\ \wedge
\xi(\var{ansn}) = 0\ \wedge
\xi(\var{ip}) = \dval{ip}\ \wedge\
\xi(\var{send\_time}) = \infty\ \wedge\\[-.22em]
\xi(\var{mqueue}) = [\hspace{.3em}]\ \wedge\
\xi(\var{pkt}) = [\hspace{.3em}]\ \wedge\
\zeta(\var{queue}) = [\hspace{.3em}]\ \\[2ex]
0 < \var{LB} < \var{LB} + \Delta\var{B} < \xi(\var{hp\_maxjitter}) < \xi(\var{hello\_interval}) < \infty\ \wedge\ \\[-.22em]
\var{LB} + 2\Delta\var{B} + \xi(\var{hello\_interval}) < \xi(\var{h\_hold\_time}) < \infty\ \wedge\ \\[-.22em]
\phantom{0 < \var{LB} <} \var{LB} + \Delta\var{B} < \xi(\var{tp\_maxjitter}) < \xi(\var{tc\_interval}) < \infty\ \wedge\ \\[-.22em]
(2(\var{LB} + \Delta\var{B}) + 1)(|\pred{IP}|-1) - (\var{LB} + 1) + \xi(\var{tc\_interval}) <
  \xi(\var{t\_hold\_time}) < \infty\ \wedge \\[-.22em]
0 \le \xi(\var{l\_hold\_time}) < \infty \\[2ex]
-\infty < \xi(\var{now}) < \infty\ \wedge\\[-.22em]
\xi(\var{now}) \le \xi(\var{hello\_time}) \le \xi(\var{now}) + \xi(\var{hello\_interval}) \ \wedge\\[-.22em]
\xi(\var{now}) \le \xi(\var{tc\_time}) \le \xi(\var{now}) + \xi(\var{tc\_interval})
\end{gather*}

\noindent
To prevent information from expiring prematurely, we require messages to be valid until another
message of the same type and from the same originating router is received. Under the assumption that
links do not change during transmission, a worst-case execution time analysis yields a maximum
elapsed time of $\var{LB} + 2\Delta\var{B}$ between HELLO messages and
$(2(\var{LB} + \Delta\var{B}) + 1)(|\pred{IP}|-1) - (\var{LB} + 1)$ between TC messages.
This explains the conditions on the constants \var{h\_hold\_time} and \var{t\_hold\_time}.

All other variables used in Processes~1--5 (\var{msgs}, \var{msg},
\var{fmprs}, \var{rmprs}, \var{rs}$'$, \var{in\_metric} and \var{x}),
as well as the variables \var{pkt}, \var{q} and \var{qs} used in Process~6, are initially undefined.

\end{document}